\documentclass[twoside,twocolumn,9pt,superscriptaddress]{revtex4-1}
\pdfoutput=1

\usepackage[left=1.5cm, right=1.5cm, top=1.785cm, bottom=2.0cm]{geometry}
\usepackage{balance}
\usepackage{mathptmx}
\usepackage{lastpage}
\usepackage{graphicx}
\usepackage{subfiles}
\usepackage{float}
\usepackage{fancyhdr}
\usepackage{fnpos}
\usepackage[english]{babel}
\addto{\captionsenglish}{%
  
}
\usepackage{array}
\usepackage{droidsans}
\usepackage{charter}
\usepackage[T1]{fontenc}
\usepackage[usenames,dvipsnames]{xcolor}
\usepackage{setspace}
\usepackage[compact]{titlesec}
\usepackage{hyperref}
\usepackage{amssymb,amsmath,amsfonts}
\usepackage{bm}
\usepackage{xfrac}
\usepackage{epstopdf}

\usepackage{tabularx}
\usepackage{multirow}
\usepackage{makecell}
\usepackage{dcolumn}

\newcommand{\eq}[1]{Eq.~\eqref{#1}}

\newcommand{\fig}[1]{Fig.~\ref{#1}}
\newcommand{\sctn}[1]{\S~\ref{#1}}
\newcommand{\tbl}[1]{Table~\ref{#1}}

\renewcommand{\vec}[1]{\mathbf{#1}}
\newcommand{\tens}[1]{\underline{\underline{#1}}}
\newcommand{\projection}[2]{\mathrm{proj}({#1};{#2})}

\newcommand{\tr}[1]{\mathrm{tr}\left[ #1 \right] }

\newcommand{\sgn}[1]{\mathrm{sgn}\left[ #1 \right] }

\newcommand{\rotfric}{\gamma_R}
\newcommand{\shearcoeff}{X}
\newcommand{\tumblingcoeff}{\lambda}

\newcommand{\rodlen}{\ell_u}
\newcommand{\mshorthand}{\mathcal{M}_b}

\newcommand{\extrapolationlen}{\xi_K}
\newcommand{\Sl}{\text{Sl}}
\newcommand{\Wr}{\text{Wr}}
\newcommand{\Tw}{\text{Tw}}
\newcommand{\basis}{\vec{e}}
\newcommand{\pos}{\vec{x}}
\newcommand{\ori}{\vec{u}}
\newcommand{\vel}{\vec{v}}
\newcommand{\surftangent}{\vec{t}}
\newcommand{\Qpos}{\vec{q}}
\newcommand{\pointcharge}{p}
\newcommand{\framing}{\mathbf{w}}
\newcommand{\angvel}{\vec{w}}

\newcommand{\impulse}{\vec{J}}
\newcommand{\angmom}{\vec{L}}
\newcommand{\dir}{\vec{n}}
\newcommand{\torque}{\boldsymbol{\Gamma}}
\newcommand{\surfnormal}{\boldsymbol{\nu}}
\newcommand{\force}{\mathbf{F}}
\newcommand{\identity}{\tens{1}}
\newcommand{\Qtens}{\tens{Q}}
\newcommand{\rotVecdG}{\bm{\Omega}}
\newcommand{\tanVecdG}{\mathbf{T}}
\newcommand{\raterotation}{\tens{W}}
\newcommand{\ratestrain}{\tens{E}}
\newcommand{\mominertia}{\tens{I}}

\begin{document}

\title{Entangled nematic disclinations using multi-particle collision dynamics}
\author{Louise C. Head}
\email{l.c.head@sms.ed.ac.uk}
\affiliation{School of Physics and Astronomy, The University of Edinburgh, Peter Guthrie Tait Road, Edinburgh, EH9 3FD, UK.}
\author{Yair A. G. Fosado}
\affiliation{School of Physics and Astronomy, The University of Edinburgh, Peter Guthrie Tait Road, Edinburgh, EH9 3FD, UK.}
\author{Davide Marenduzzo}
\affiliation{School of Physics and Astronomy, The University of Edinburgh, Peter Guthrie Tait Road, Edinburgh, EH9 3FD, UK.}
\author{Tyler N. Shendruk}
\email{t.shendruk@ed.ac.uk}
\affiliation{School of Physics and Astronomy, The University of Edinburgh, Peter Guthrie Tait Road, Edinburgh, EH9 3FD, UK.}

\begin{abstract}

Colloids dispersed in nematic liquid crystals form topological composites in which colloid-associated defects mediate interactions while adhering to fundamental topological constraints. 
Better realising the promise of such materials requires numerical methods that model nematic inclusions in dynamic and complex scenarios. 
We employ a mesoscale approach for simulating colloids as mobile surfaces embedded in a fluctuating nematohydrodynamic medium to 
study the kinetics of colloidal entanglement. 
In addition to reproducing far-field interactions, topological properties of disclination loops are resolved to reveal their metastable states and topological transitions during relaxation towards ground state.
The intrinsic hydrodynamic fluctuations distinguish formerly unexplored far-from-equilibrium disclination states, including configurations with localised positive winding profiles. 
The adaptability and precision of this numerical approach offers promising avenues for studying the dynamics of colloids and topological defects in designed and out-of-equilibrium situations.
\end{abstract} 

\maketitle
\newpage

\section{Introduction}

Dispersions of colloidal particles in liquid crystals~\cite{stark2001} are of interest to physicists because they provide a pathway to realise soft materials with interesting target properties, such as photonic crystals~\cite{smalyukh2018}, cloaks and metamaterials~\cite{lavrentovich2011}, or self-quenched glasses~\cite{wood2011}. This versatility is due to the fact that topology and elastic distortions in the liquid crystalline host lead to long-range interactions which can be tuned by varying particle size, shape and liquid crystalline properties, even in a simple nematic. When combined with a suitable kinetic protocol, these interactions can be harnessed to self-assemble different types of materials~\cite{musevic2006}.

To understand the physical mechanisms underlying the self-assembly of different structures, a useful and popular starting point is that of two colloidal particles in a nematic, with normal anchoring at the colloidal surface. On the one hand, analysing this geometry leads to an estimate of the effective pair potential between particles, which includes elasticity and defect-mediated interactions, and which is important for self-assembly in many-particle systems~\cite{copar2011n1,copar2011n2,tasinkevych2002}. On the other hand, the problem of a colloidal dimer in a liquid crystal is interesting from a fundamental point of view, due to the central role played by topology~\cite{alexander2012}. Indeed, the liquid crystalline pattern needs to be topologically trivial overall~\cite{copar2011n1,copar2011n2}, but this can be realised in a number of possible ways. For instance, each colloid can be surrounded by a topologically charged Saturn ring~\cite{terentjev1995,gu2000}, as the total topological charge in the system only needs to equal $0$ modulo $2$ in three dimensions~\cite{mermin1979,copar2011n1}. However, another topologically allowed configuration is one where a single writhed disclination loop wraps around both colloids. Configurations such as these are referred to as {\it entangled disclinations}, and the writhe in the loop cancels the topological charge which would otherwise be present \cite{copar2011n2}. The relation between writhe and topological charge can be understood by introducing the self-linking number~\cite{copar2011n1,copar2014}, which describes the topology of a disclination loop, in the case where the local director field profile (in the plane perpendicular to the loop tangent) is topologically equivalent to that of a planar defect with winding number $-1/2$, or a triradius. In such cases, the loop possesses the same topology as a ribbon~\cite{copar2011n2}. 
In this way, colloids dispersed in liquid crystals can act as probes for fundamental questions of topology. 

Colloidal dispersions in nematics have mainly been studied with continuum models, either via free energy minimisation techniques~\cite{copar2011n2,wood2011}, or by means of hybrid lattice Boltzmann simulations~\cite{lintuvuori2011}. In this work, we employ a different methodology to study a single colloid or a pair of colloids in a nematic host, based on multi-particle collision dynamics (MPCD).
Though it was traditionally applied to moderate-P\'{e}clet number situations within isotropic fluids, the MPCD algorithm has recently been extended to simulate fluctuating, linear nematohydrodynamics~\cite{shendruk2015,Hijar2019} or to be hybridized with continuum descriptions of the nematic~\cite{lee2015,lee2017,Mandal2019,mandal2021}. 
Importantly, this nematic algorithm (N-MPCD) captures the competing influences of thermal fluctuations, elastic interactions, and hydrodynamics, and hence can be used to study the topological evolution of defect structures over time. 
The natural inclusion of noise makes it possible to consider the case of small particles, where the free energy profile of the system is rid of large barriers, which otherwise dominate the colloidal kinetics~\cite{lavrentovich2011}. The fact that N-MPCD provides a particle-based description of the nematic fluid also simplifies the treatment of boundary conditions, and hence makes it easier to extend this algorithm to complex surface geometries, such as rodlike particles~\cite{hijar2020} or wavy channels~\cite{Wamsler2024}. Additionally, MPCD can be readily extended to study active nematics~\cite{kozhukhov2022,macias2023} and systems with many colloids, thereby providing a powerful package to study the hydrodynamics of topological composite materials~\cite{senyuk2013}. 

Here, the N-MPCD algorithm is validated by computing the elastic force between a colloid and a wall, or between two colloids. These follow scaling laws in agreement with previous theoretical predictions and numerical estimates. The topological patterns are studied, both over time and in steady state with a single colloidal particle or a colloidal dimer. The steady-state patterns broadly confirm the set of structures predicted in the literature by elastic energy minimisation~\cite{ravnik2009colloids}. 
Thus, a pair of Boojums for colloids with tangential anchoring are found. 
With normal anchoring, a Saturn ring and a dipolar halo are found. A colloidal dimer with normal anchoring results in either two topologically charged loops or an uncharged but writhed loop with non-trivial self-linking numbers. However, thermal fluctuations and boundary influences can lead to tilted and non-ideal versions of these entangled structures. 
Although disclination loops are always associated with local director field patterns with $-1/2$ profiles in steady state, a wider variety of states are observed en route to equilibrium. These are found to differ substantially in their geometric features. 
Examples of transient patterns include longer loops with twist and even $+1/2$ local director profiles, as well as skewed rings. 

\section{Methods}

Multi-Particle Collision Dynamics is a coarse-grained mesoscopic particle-based hydrodynamic solver that is versatile for simulating a wide variety of Newtonian~\cite{malevanets1999}, complex~\cite{winkler2005} and active fluids~\cite{kozhukhov2022,macias2023}. 
It has found particular utility simulating suspensions of polymers~\cite{lamura2001}, colloids~\cite{ReyesArango2020,shendruk2013} and bacteria~\cite{zantop2021,zottl2018}, because its intrinsic thermal noise makes it ideal for moderate P\'{e}clet numbers. 
Since N-MPCD can support elastic and hydrodynamic interactions, combined with thermal diffusivity, this offers a promising avenue for studying topological microfluidics~\cite{sengupta2015,copar2021}, design principles for self assembly kinetics, defect interactions with active fields~\cite{kozhukhov2022,macias2023}, and microfluidic transport through harnessing energy landscapes~\cite{Wamsler2024,luo2018}.

\subsection{Bulk nematohydrodynamic evolution}

Nematic Multi-Particle Collision Dynamics discretises continuous hydrodynamic fields for mass, momentum and orientational order into $N$ point particles, indexed by $i$, with mass $m_i$, position $\pos_i$, velocity $\vel_i$ and orientation $\ori_i$ in $d$ dimensions. 
N-MPCD is a two-step algorithm, in which particles evolve through \textit{(i)} streaming and \textit{(ii)} collision steps, dictating how the particles move and interact with their local environment~\cite{malevanets1999}. 

The streaming step controls the spatial evolution of each particle position, defined as ballistic streaming over the time interval $\delta t$
\begin{align}
    \label{eq:streaming}
    \pos_i(t+\delta t) &= \pos_i(t) + \vel_i(t) \delta t.
\end{align}
The collision step represents inter-particle interactions that have been coarse-grained into a lattice of cells, indexed by $c$, each containing $N_c$ particles. Particles interact only with their local cell environment via collision operators, which avoid the demanding computational cost of explicitly calculating all pair-wise interactions, and are shown to reproduce hydrodynamic fields over sufficiently long length- and timescales. 
Hydrodynamic-scale fields are extracted through cell-based averaging, $\phi_{c}(t) = \langle \phi_i\rangle_c = \sum_{i}^{N_c}\phi_{i}(t) / N_{c}$. 
The evolution equations for $\vel_i$ and $\ori_i$ have contributions from cell-based momentum-conserving collision operators. 
First considering the translational momentum collision
\begin{align}
    \label{eq:velcollision}
    \vel_i(t+\delta t) &= \vel_{c}(t) + \boldsymbol{\Xi}^\mathrm{vel}_{i,c}(t). 
\end{align}
The collision operator $\boldsymbol{\Xi}^\mathrm{vel}_{i,c}(t) = \boldsymbol{\Xi}^\mathrm{vel,iso}_{i,c} + \boldsymbol{\Xi}^\mathrm{vel,nem}_{i,c}$ has two contributions: an isotropic part $\boldsymbol{\Xi}^\mathrm{vel,iso}_{i,c}$, and a nematic backflow contribution $\boldsymbol{\Xi}^\mathrm{vel,nem}_{i,c}$, the latter of which will be discussed after the orientation contributions. The isotropic collision uses the Andersen locally thermostatted collision operator~\cite{Gompper2007EPL,
Gompper2007PRE}
\begin{align}
   \boldsymbol{\Xi}i^\mathrm{vel,iso}_{i,c} &= \boldsymbol{\xi}_i - \boldsymbol{\xi}_c + (\mominertia^{-1}\cdot \delta \angmom_{\mathrm{vel}})\times \pos_i' ,
\end{align}
where $\boldsymbol{\xi}_i$ are randomly generated from a Gaussian distribution with variance $k_\mathrm{B} T/m$, and $\boldsymbol{\xi}_c = \langle \boldsymbol{\xi}_i\rangle_c$ is a residual term, designed to conserve the net linear momentum from the noise. The third term is a correction to conserve angular momentum, for particles located about the center of mass $\pos_i'=\pos_i-\pos_c$ with a moment of inertia tensor $\mominertia$ and angular momentum $\angmom_{\mathrm{vel}}$ about $\pos_c$.
Since the collision operator is applied to lattice-based cells, a random grid shift is included to preserve Galilean invariance~\cite{ihle2001,gompper2009}.

A cell-based collision operation is also applied to orientations
\begin{align}
    \label{eq:oricollision}
    \ori_i(t+\delta t) &= \dir_{c}(t) + \Xi^{\mathrm{ori}}_{i,c}(t) .
\end{align}
about the cell's local director $\dir_{c}(t)$. Constructing a cell-based nematic tensor order parameter, $\Qtens_c=\frac{1}{d-1}\langle d\ori_i \ori_i - \identity\rangle_c$, allows the local scalar order parameter $S_c$ and director $\dir_c$ to be found as the largest eigenvalue and corresponding eigenvector. Treating the cell's orientational order parameters as a mean field, the orientation collision $\Xi^{\mathrm{ori}}_{i,c}$ stochastically draws orientations from a local Maier-Saupe distribution $f_{\mathrm{ori}}= f_0 \exp\left(U S_c (\ori_i \cdot \dir_c)^2 / k_\mathrm{B} T\right)$, centered about $\dir_c$ with a normalisation constant $f_0$ and a mean field interaction constant $U$. 
The interaction constant is linearly proportional to the one-constant approximation of Frank elasticity $K$~\cite{shendruk2015}. For large $U$, the particle orientations are deep in the nematic phase, aligning close to the free energy minimum, with small thermal fluctuations. 

Nematohydrodynamics requires coupling terms in \eq{eq:velcollision} and \eq{eq:oricollision} to account for velocity gradients rotating orientations and orientational motion generating nematic backflows.  
This can be cast in terms of an overdamped bulk-fluid torque equation for each particle $i$
\begin{align}
    \label{eq:torquebalance}
    \torque^{\mathrm{col}}_i + \torque^{\mathrm{HI}}_i + \torque^{\mathrm{diss}}_i &= 0 . 
\end{align}
The torques from the orientational collision ($\torque^\mathrm{col}$) and hydrodynamic flows ($\torque^\mathrm{HI}$) can be written as $\torque^{\mathrm{col}}_i + \torque^{\mathrm{HI}}_i = \gamma_R \ori_i \times \left(\sfrac{\delta \ori_i^\mathrm{col}}{\delta t} + \sfrac{\delta \ori_i^{\mathrm{HI}}}{\delta t}\right)$, where $\gamma_R$ is a rotational friction coefficient. 
From \eq{eq:oricollision}, the collisional contribution is $\sfrac{\delta \ori_i^\mathrm{col}}{\delta t} = \sfrac{\left(\dir_c(t) + \Xi_{i,c}^{\mathrm{ori}}(t) \right)}{\delta t}$. 
The hydrodynamic contribution applies Jeffery coupling between the orientation and velocity gradient, $\sfrac{\delta \ori^{\mathrm{HI}}}{\delta t} = \shearcoeff \left[\vel_i \raterotation + \tumblingcoeff \left(\ori_i \cdot \ratestrain - \ori_i\ori_i\ori_i : \ratestrain \right) \right]$, where $\shearcoeff$ is a shear coupling coefficient that influences the relaxation time of alignment relative to $\delta t$, $\tumblingcoeff$ is the flow tumbling parameter, and $\ratestrain$ and $\raterotation$ are the symmetric and skew-symmetric components of the velocity gradient tensor. 
The remaining contribution is the dissipative torque $\torque^\mathrm{diss}$, which is converted into backflow in the velocity evolution equation through an angular momentum correction $\Xi^\mathrm{vel,nem}_{i,c} = -\mominertia^{-1} \cdot \delta \angmom_\mathrm{ori} \times \pos_i'$, where $\delta\angmom_{\mathrm{ori}} = \sum_{i}^{N_c}\torque^\mathrm{diss}_{i}\delta t$, which goes into \eq{eq:velcollision}.

\subsection{Boundary conditions}
\label{sctn:BCs}

\begin{figure*}[tb]
    \centering
    \includegraphics[width=0.9\linewidth]{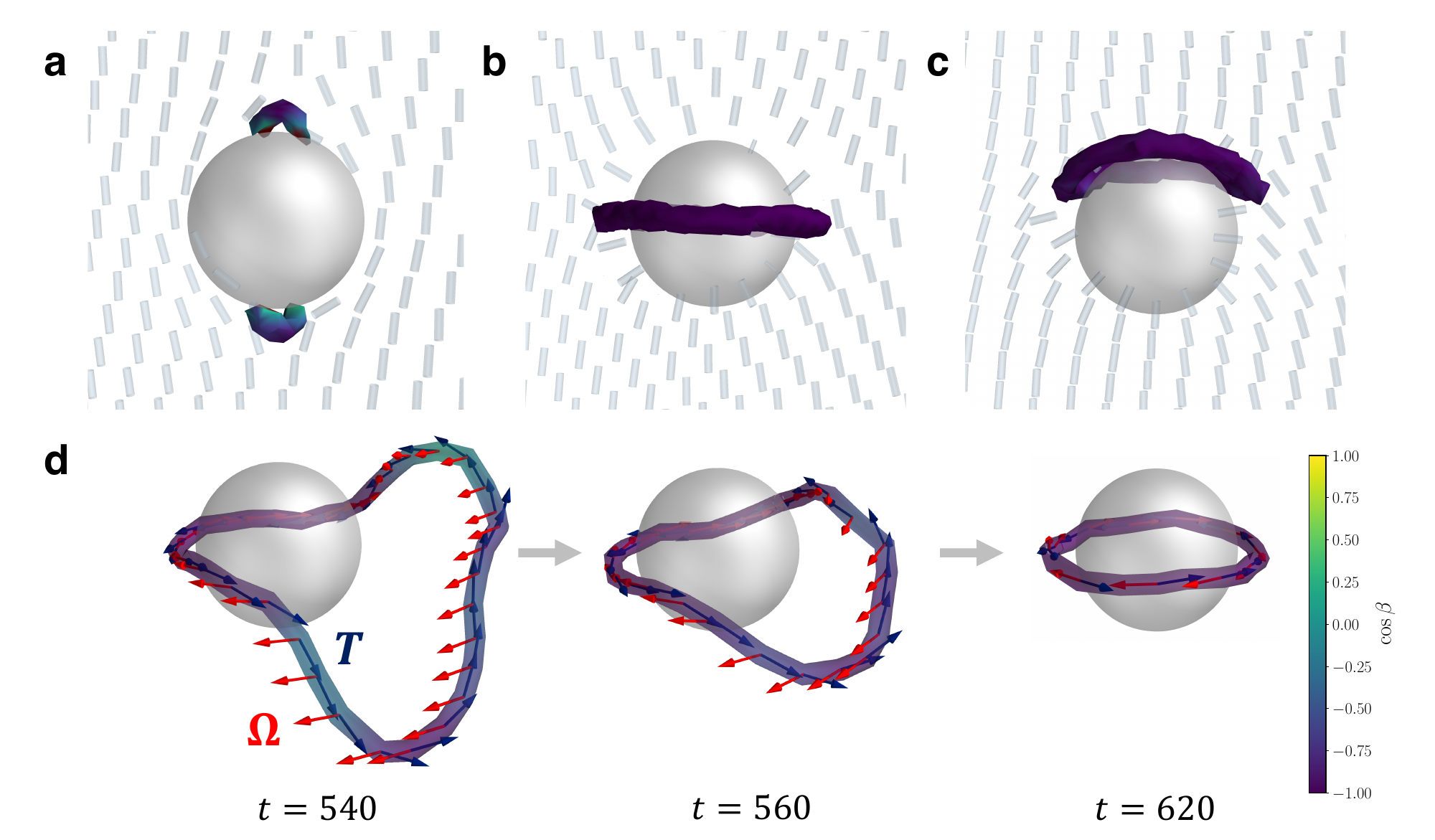}
    \caption{Steady-state defect configurations associated with a single colloid in N-MPCD. 
    \textbf{(a)} Boojum defects. 
    \textbf{(b)} Saturn ring. 
    \textbf{(c)} Hyperbolic hedgehog opened into a halo-like ring. 
    \textbf{(d)} Dynamics of a near-colloid disclination loop relaxing into a Saturn ring. 
    The disclination rotation vector $\rotVecdG$ and tangent vector $\tanVecdG$ are shown as red and navy blue arrows. 
    Each frame is shown with the corresponding simulation time $t$. 
    In \textbf{(a)}-\textbf{(d)}, each defect is visualised as an isosurface of $s(\pos)=0.9$ (\sctn{sctn:defects}). 
    The disclinations are coloured by $\cos\beta=\rotVecdG\cdot\tanVecdG$. Purple represents a $-1/2$ wedge profile, yellow a $+1/2$ wedge profile and green twist type.
    }
    \label{fig:singleColloid}
\end{figure*}

The bulk fluid domain is maintained by \textit{(i)} defining surface equations representing boundaries, and \textit{(ii)} setting rules on N-MPCD particles that violate the surface equation. Each boundary, with index $b$, has a surface equation with an implicit form $S_b(\pos)=0$, where $\pos$ satisfy the set of points on the surface. Particles violate a surface equation if
\begin{align}
    \label{eq:surfviolation}
    S_b(\pos_i) &\leq 0,
\end{align}
corresponding to particles streaming inside. Particles are ray-traced back to the surface boundary at position $\pos_i^*$, at time $t^*<\delta t$ (found where particle path and $S_b(\pos_i)=0$ intersect). Boundary rules are then applied, and the particle resumes streaming for remaining time $\delta t - t^*$.

Boundary rules operate on the particle's generalised coordinates, $\pos_i$, $\vel_i$ and $\ori_i$. For periodic boundary conditions, $\pos_i\rightarrow \pos_i + D_{\surfnormal_b} \surfnormal_b$ where $D_{\surfnormal_b}$ is a scalar shift in the surface normal direction $\surfnormal_b$ of boundary $b$. Operators on the velocity are required for solid impermeable walls, $\vel_i \rightarrow M_{\surfnormal_b} \projection{\vel_i}{\surfnormal_b} + M_{\surftangent_b} \projection{\vel_i}{\surftangent_b}$, where $M_{\surfnormal_b}$ and $M_{\surftangent_b}$ are scalar multipliers on the projection of $\vel_i$ in the surface normal $\surfnormal_b$ and tangent $\surftangent_b$ directions.
The surface normal projections have the form $\projection{\vec{f}}{\surfnormal}=(\surfnormal \cdot \vec{f})\surfnormal$ and surface tangent projections, $\projection{\vec{f}}{\surftangent}=\vec{f} - (\surfnormal \cdot \vec{f})\surfnormal$. 
No-slip boundary conditions require bounce-back multipliers $M_{\surfnormal_b}=M_{\surftangent_b}=-1$. Additionally ghost particles are required to ensure that $\vel_c$ extrapolates to zero in cells that are intersected by boundaries~\cite{lamura2001,lamura2002}. Anchoring conditions operate on the particle's orientation through $\ori_i \rightarrow \mathcal{M}_{\surfnormal_b} \projection{\ori_i}{\surfnormal_b} + \mathcal{M}_{\surftangent_b} \projection{\ori_i}{\surftangent_b}$, with the constraint that $\ori_i$ maintains unit magnitude ($\ori_i \cdot \ori_i = 1$). Homeotropic (normal) anchoring is achieved with $\mathcal{M}_{\surfnormal_b}=1$ and $\mathcal{M}_{\surftangent_b}=0$, and planar (tangential) anchoring with $\mathcal{M}_{\surfnormal_b}=0$ and $\mathcal{M}_{\surftangent_b}=1$.

Despite $\mathcal{M}_{\surfnormal_b}$ and $\mathcal{M}_{\surftangent_b}$ setting the orientation of any particles that violate \eq{eq:surfviolation}, the anchoring is not infinitely strong. 
This is because, of the $N_c$ particles in any cell that intersects $S_b$ only some fraction $N^*_c/N_c$ would have collided with the surface. 
Although those $N^*_c$ particles have their orientation set, the collision operation (\eq{eq:oricollision}) stochastically exchanges orientations between all $N_c$ particles, effectively weakening the anchoring condition. 
To strengthen the anchoring, the orientational boundary condition is applied to all $N_c$ particles within cells that are intersected by the surface $S_b$ (\sctn{appendix:klemen-deGennes}).

\subsection{Mobile colloids}
\label{section:mobilecolloids}

One way to incorporate colloids is to include them as embedded molecular dynamics particles, with radial interaction potentials~\cite{hijar2020,ReyesArango2020}.
In contrast, the present work treats each colloid as a mobile surface that interacts with the hydrodynamic fields via conserving the linear and angular impulse generated by each of the incremental particle transformations. 
The surface equation
\begin{align}
    \label{eq:sphere}
    S_b(\pos) &= \left[ \pos - \Qpos_b(t) \right]^2 - R^2 = 0,
\end{align}
defines spherical colloids featuring a temporally-varying centre coordinate $\Qpos_b(t)$ and constant radius $R$. 

Analogous to the particle streaming \eq{eq:streaming}, the colloid coordinate translates assuming ballistic streaming $\Qpos_b(t+\delta t)= \Qpos_b(t) + \vel_b(t) \delta t$, where $\vel_b(t)$ is the colloid's centre of mass velocity, which is sufficient under the viscously overdamped assumption. Since spheres have inherent rotational symmetry, \eq{eq:sphere} is invariant under colloid rotation with angular velocity $\angvel_b$, defined relative to $\Qpos_b$. 
Each colloidal $\vel_b(t)$ and $\angvel_b(t)$ are determined by the incremental sum over all $N_b^*$ particles that violate \eq{eq:surfviolation} in the current timestep
\begin{align}
    \label{eq:velcolloid}
    \vel_b(t+\delta t) &= \vel_b(t) + \sum_i^{N_b^*}  \delta \vel_{b,i}^{\mathrm{vel}} + \sum_i^{N_c}\delta\vel_{b,i}^{\mathrm{ori}}\\ 
    \label{eq:angvelcolloid}
    \angvel_b(t+\delta t) &= \angvel_b(t) + \sum_i^{N_b^*} \delta \angvel_{b,i}^{\mathrm{vel}} + \sum_i^{N_c}\delta\angvel_{b,i}^{\mathrm{ori}} ,
\end{align}
where $\mathrm{vel}$ superscript corresponds to changes from the velocity boundary conditions, and $\mathrm{ori}$, from the orientation rules. The orientation contributions sum over all $N_c$ particles within cells that intersect a colloid boundary (\sctn{appendix:klemen-deGennes}).
The contributions from velocity rules, enter as an impulse created by the change in momentum of the particle's velocity $\impulse_i = m_i\vel_i(t+\delta t) - m_i\vel_i(t)$~\cite{shendruk2013}. Balancing by an impulse on the colloid $\impulse_b=-\impulse_i$ leads to
\begin{align}
    \label{eq:veltoboundary}
    \delta \vel_{b,i}^{\mathrm{vel}} &= \projection{\impulse_b}{\surfnormal_b}/m_b \\
    \delta \angvel_{b,i}^{\mathrm{vel}} &= \mominertia_b^{-1} \cdot \left(\vec{r}_{b,i}\times\projection{\impulse_b}{\surftangent_b}\right) ,
\end{align}
where $m_b$ is the mass of the colloid, $\mominertia_b$ is the moment of inertia and $\vec{r}_{b,i}$ is the vector from the centre of the colloid to the collision point on the boundary.
The contributions from orientation rules are calculated from conserving a torque balance due to anchoring 
\begin{align}
    \label{eq:anchtorquebalance}
    \torque_i^{\mathrm{anch}} + \torque_{b,i}^{\mathrm{anch}} &= 0,
\end{align}
where $\torque_i^{\mathrm{anch}}$ corresponds to the particle reorientation to prescribed anchoring condition and $\torque_{b,i}^{\mathrm{anch}}$ is the torque felt by the boundary to balance the particle reorientation event. The anchoring torque to align either with homeotropic or planar anchoring can be written in terms of the initial orientation and surface normal
\begin{align}
    \label{eq:anchoringtorque}
    \torque_i^{\mathrm{anch}} &= \frac{\rotfric}{\delta t}\mshorthand(\ori_i \cdot \surfnormal_b)(\ori_i \times \surfnormal_b),
\end{align}
where $\mshorthand = \left( \mathcal{M}_{\surfnormal_b}-\mathcal{M}_{\surftangent_b} \right) \left( \mathcal{M}_{\surfnormal_b}^2+\mathcal{M}_{\surftangent_b}^2 \right)^{-1/2} = +1$
for homeotropic, and $\mshorthand=-1$ for planar anchoring (\sctn{SI:anchoringtorque}). The denominator ensures that the final particle orientation has unit magnitude. By defining the angle $\cos\alpha_i=\ori_i\cdot \surfnormal_b$, the torque magnitude can be written in terms of a single variable $\torque_i^{\mathrm{anch}} = \frac{\rotfric}{2 \delta t}\mshorthand\sin 2\alpha_i$.
The odd nature of $\torque_i^{\mathrm{anch}}$ with respect to $\alpha_i$, means that the torque balance can be satisfied by introducing a virtual particle, oriented initially at $-\alpha_i$ to $\surfnormal_b$ (with orientation unit vector $\ori_{b,i}$). Over the time $\delta t$, the virtual particle reorients to align with $\surfnormal_b$ through application of the torque $-\torque_i^{\mathrm{anch}}$. 
The initial orientation of the virtual particle $\ori_{b,i}$ is related to the N-MPCD particle $\ori_i$ by a mirror reflection about $\surfnormal_b$.

Torque is converted to a force acting on the boundary via 
\begin{align}
    \force_{b,i}^{\mathrm{anch}} = \frac{ \torque_{b,i}^{\mathrm{anch}}\times \ori_{b,i}}{ \rodlen/2},
    \label{eq:torque2force}
\end{align}
neglecting the colinear terms (\sctn{SI:torquetoforce}). In determining the rotation effect, $\ell_u$ is required to represent the lengthscale of the MPCD nematogens and control the rotational susceptibility.
The head-tail symmetry of the particle orientation $\ori_{b,i}$ provides ambiguity on the sign of $\force_{b,i}^{\mathrm{anch}}$, 
which is chosen to be oriented towards the boundary as $\force_{b,i}^{\mathrm{anch}} \cdot \surfnormal_b < 0$. 
For spherical colloids, the force at the boundary can be converted into linear and angular velocity contributions, through projecting $\force_{b,i}^{\mathrm{anch}}$ in the surface normal and tangential directions
\begin{align}
    \label{eq:oritoboundaryvel}
    \delta \vel_{b,i}^{\mathrm{ori}} &= \projection{\force_{b,i}}{\surfnormal_b}\delta t/m_b \\
    \label{eq:oritoboundaryangvel}
    \delta \angvel_{b,i}^{\mathrm{ori}} &= \mominertia_b^{-1} \cdot  \left( \vec{r}_{b,i}\times\projection{\force_{b,i}}{\surftangent_b}\delta t \right). 
\end{align}

\subsection{Units and parameters}
\label{sctn:unitsandparam}

Values are given in MPCD units of cell size $a=1$, particle mass $m=1$ and thermal energy $k_\mathrm{B} T=1$. 
This results in units of time $\tau = a\sqrt{m/k_\mathrm{B} T} = 1$. 
Simulation time iterates with time-step size $\delta t = 0.1$. Simulations are performed in two ($d=2$) and three ($d=3$) dimensions with system sizes $[L_x,L_y]$ and $[L_x,L_y,L_z]$ respectively, aligned with a Cartesian basis $\basis_x,\basis_y,\basis_z$. The average particle density per cell is $\langle N_c \rangle = 20$.  
The nematic mean field potential is set to $U=20$, corresponding to deep in the nematic phase~\cite{shendruk2015}. Other nematohydrodynamic parameters include the rotational friction $\rotfric = 0.01$, shear susceptibility $\shearcoeff=0.5$ and tumbling parameter set to be in the shear aligning regime with $\tumblingcoeff=2$.
Unless otherwise stated, colloids with radii $R=6$ are used in three-dimensions, and $R=10$ in two-dimensions.
The effective particle rod-length $\ell_u=0.006$, tunes the strength of the interaction between nematic bulk elasticity and colloid mobility. In all simulations, MPCD particles start with randomly generated positions and velocities.
While the bulk fluid properties remain constant between simulations, the boundary conditions vary between studies, in addition to initial particle orientations. 
Additional system specific parameters are given in the Appendix. 

\section{Results}

\subsection{Defects around a single colloid}

\begin{figure*}[tb]
    \centering
    \includegraphics[width=0.9\linewidth]{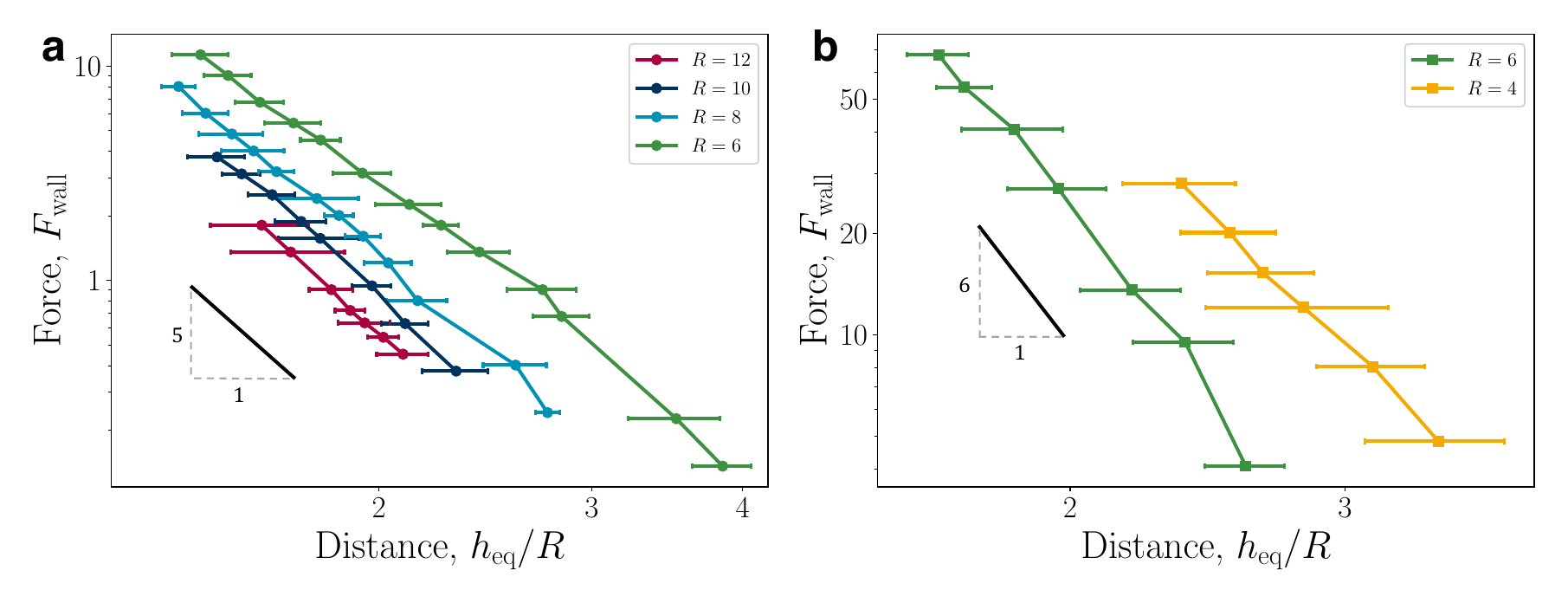}
    \caption{
    Elastic interaction forces between quadrupolar colloids of radius $R$ and a homeotropic anchored wall. \textbf{(a)} Two-dimensional elastic repulsion force $F_\mathrm{wall}$ measured at equilibrium colloid-wall separation distances $h_{\mathrm{eq}}$. Distances are scaled by the colloid radius $R$. \textbf{(b)} Same as \textbf{(a)} but for three-dimensions. The black lines, shown for comparison with the expected scaling (\eq{eq:colloidWall}), represent decaying power laws of -5 (2D) and -6 (3D). 
    Errorbars represent the standard error between independent runs.}
    \label{fig:forceDist}
\end{figure*}

To examine the defect structures around isolated nematic colloids, a single sphere is initialised within a thermal quench (randomised orientations) in an $L_x=L_y=L_z=40$ domain with periodic boundary conditions on all walls. The simulations are
run for a duration of $T_S = 1400$, with data recorded for $t\geq400$.
After long times ($t\sim 600$), the nematic field approaches its equilibrium state, which includes static defects that accompany the colloidal particle (\fig{fig:singleColloid}a-c). 
For the case of planar anchoring (\fig{fig:singleColloid}a), the inability for a tangential vector field to continuously coat a sphere necessitates two surface defects at the colloidal antipodes, known as Boojums \cite{mermin1990,Liu2013}. 
Their two opposite surface defects give the colloid/Boojums complex a quadrupolar structure.
For spherical surfaces, these can either be hyperbolic point defects split in half by the mirror plane of the colloid, or separated into handle-shaped semi-loops that connect two $+1/2$ closely separated surface defects~\cite{Liu2013}, with the latter case being observed for simulations from a quench (\fig{fig:singleColloid}a). 

Colloids with homeotropic anchoring supply the bulk fluid with a hedgehog charge (point charge) of $\pointcharge=1$ (\fig{fig:singleColloid}b-c). 
This nucleates one of two configurations, each of which has an odd point charge to conserve topological charge.
The first configuration is a Saturn ring --- a closed $-1/2$ disclination loop surrounding the equatorial axis~\cite{terentjev1995,gu2000} (\fig{fig:singleColloid}b). 
The Saturn ring results in a quadrupolar far-field character.
The second configuration is a hyperbolic hedgehog, forming a topological dipole with the colloid~\cite{poulin1997}, which in N-MPCD manifests as a dipolar halo (\fig{fig:singleColloid}c). 
Of the $20$ independent simulations, $17$ ended with a Saturn ring, and $3$ with a dipolar halo.
In experiments, topological dipoles are the stable state when the ratio of colloid radius to Kleman-de~Gennes extrapolation length 
$R/\xi$ is large (see \sctn{appendix:klemen-deGennes}), while Saturn rings are preferred in confinement and for smaller colloids with weaker anchoring (larger extrapolation length)~\cite{stark2001,Kos2019}. 
Generally, simulations predominantly reproduce Saturn rings~\cite{andrienko2001,ruhwandl1997num2} and this is shown to be true in N-MPCD as well. For the three dimensional colloids considered here, $R/\xi \approx 40$ (\sctn{appendix:klemen-deGennes}). 

As a fluctuating nematohydrodynamic solver, N-MPCD can also simulate the coarsening dynamics of the disclination loops (\fig{fig:singleColloid}d). 
Soon after the quench, the nematic field far from the colloid has ordered, but a single, large loop remains, relaxing into a Saturn ring configuration.
The loop is free to sample disclination profiles outside of purely trefoil-like $-1/2$. This is demonstrated by colouring the disclinations with $\cos\beta=\rotVecdG\cdot\tanVecdG$ where $\rotVecdG$ is the rotation vector~\cite{friedel1969}
and $\tanVecdG$ is the tangent vector of the line. 
Where $\cos\beta=1$, $\rotVecdG$ is parallel to $\tanVecdG$ and the disclination line has a local $+1/2$ wedge profile. 
On the other hand, where $\cos\beta=-1$, $\rotVecdG$ is antiparallel to $\tanVecdG$ and the disclination locally has a $-1/2$ wedge profile. 
The director can also rotate out of this plane passing through $\cos\beta=0$, which represent twist-type profiles. 
Visualising disclinations in this way has been particularly insightful for interpreting disclination behaviours during phase transitions \cite{velez2021} and in three-dimensional active nematics~\cite{duclos2020,shendruk2018,negro2024}.  
The loop in \fig{fig:singleColloid}d is charged, requiring $\rotVecdG$ to make a full revolution. 
However, the rotation is not homogeneous and $\rotVecdG$ remains largely uniform for large segments of the disclination that are distant from the colloid. 
Conversely, the segments of the disclination closest to the colloidal surface support nearly the entire variation of $\rotVecdG$.  
At later times ($t \sim 600$), the loop reduces in size and the anchoring constraint on the colloid enforces $\rotVecdG$ to rotate into the expected anti-parallel configuration $\rotVecdG\cdot \tanVecdG=-1$, forming the Saturn ring. 
 
\subsection{Elastic interactions}

\begin{figure}[tb]
    \centering
    \includegraphics[width=\linewidth]{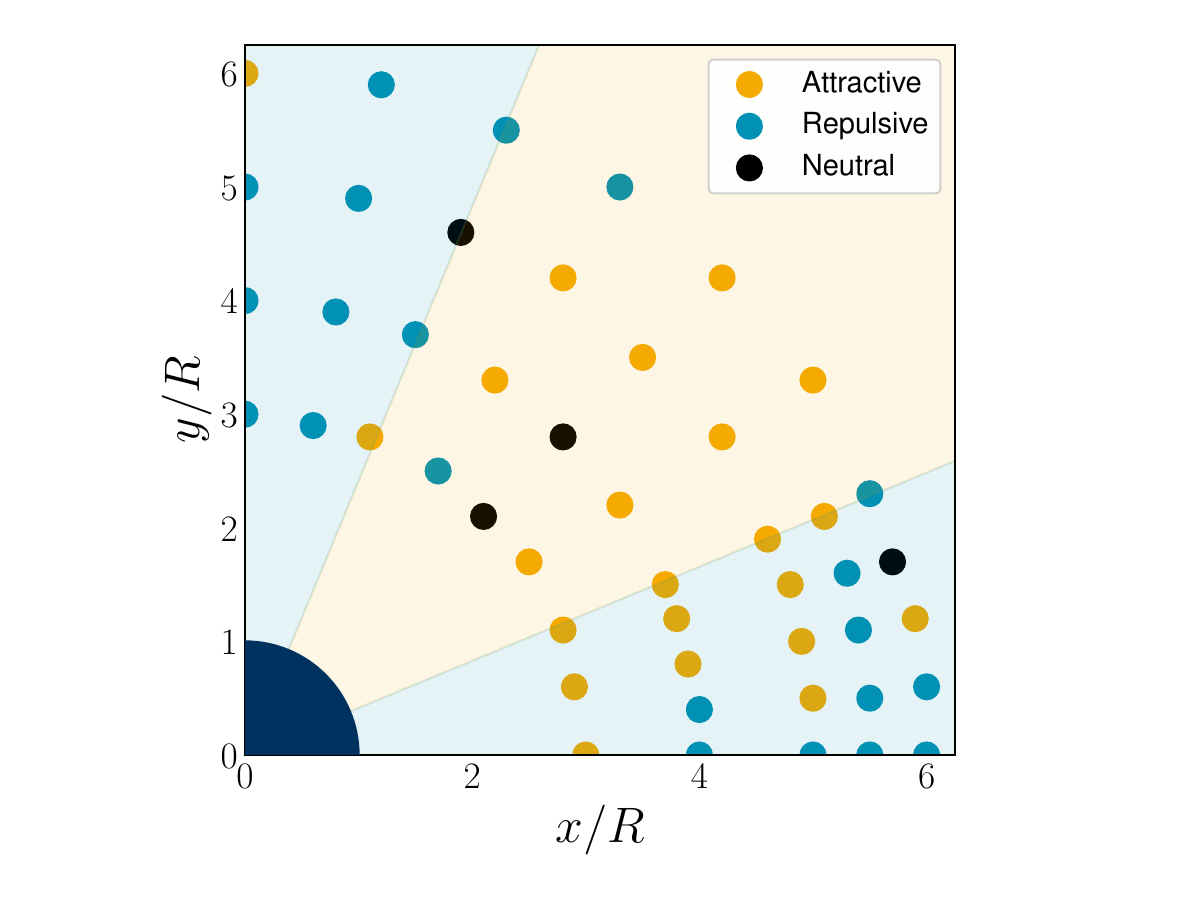}
    \caption{
    Attraction-repulsion zones between two-dimensional interacting quadrupolar colloids with radius $R=10$. 
    One colloid is fixed at the origin (navy blue quarter circle) and a second mobile colloid probes the interaction at surrounding points. 
    The colour of each point represents attractive (orange), repulsive (blue) or neutral (black) interactions. 
    Background shading shows the far-field expectation for interacting quadrupoles (\eq{eq:colloidPair}).}
    \label{fig:anisotropy}
\end{figure}

Colloid-defect complexes with homeotropic anchoring can have a quadrupolar (Saturn ring;  \fig{fig:singleColloid}b) or dipolar (dipolar halo; \fig{fig:singleColloid}c) nature~\cite{lubensky1998}. 
These configurations correspond directly to the form of far-field elastic interactions between pairs of nematic colloids. 
N-MPCD reproduces elastic forces that are long ranged, with power laws dictated by the dominant multipole moment (\sctn{sctn:power-law}), as well as anisotropic, with attraction and repulsion zones with angular variation between interacting colloids (\sctn{sctn:aniInt}).

\subsubsection{Power-law forces}
\label{sctn:power-law}

To quantify the power-law nature of nematic interactions in N-MPCD, a colloid interacting with a wall with strong homeotropic anchoring is considered. This setup is preferred over a pair of mobile colloids because it removes additional complexities arising from the relative orientation of a pair of nematic colloids. 
In the proximity of the wall, the colloid experiences a strong elastic repulsive force, $\force_\mathrm{wall}$, that decays with distance $h$~\cite{chernyshuk2011,pergamenshchik2009}. This can be represented as a quadrupole-quadrupole interaction between the colloid and its mirror image
on the other side of the wall~\cite{chernyshuk2011,dolganov2006,muller2020,tasinkevych2002} 
\begin{align}
    \mathbf{F}_\text{wall} &\sim K\frac{R^{2d}}{h^{d+3}}\surfnormal_\mathrm{wall},
    \label{eq:colloidWall}
\end{align}
in $d$ dimensions and $\surfnormal_\mathrm{wall}$ is oriented normal to the wall.  
For determining the repulsive elastic force between a homeotropic-anchored colloid and a homeotropic-anchored wall, measurements are performed in both two and three dimensions. 
A constant (gravitational-like) force $\vec{F}_\mathrm{G}$ is applied to the colloid, pushing it towards the anchored wall. 
This acts as a probe of the strength of the elastic force via the resulting equilibrium height $h_\mathrm{eq}$ that results from the balance with elastic repulsion (simulation details provided in \sctn{sctn:wallSI}). 

Elastic forces are largest at smaller colloid separations from the wall, with magnitudes $F_\text{wall}\approx10$ for $h_{\mathrm{eq}}/R\approx 1.5$ in 2D (\fig{fig:forceDist}a) and $F_\text{wall}\approx50$ for $h_{\mathrm{eq}}/R\approx 1.6$ in 3D (\fig{fig:forceDist}b). At increasing
separations, these forces rapidly decay. Comparing with predictions (\eq{eq:colloidWall}; black slope), N-MPCD elastic forces decay with the expected power laws. In two-dimensions, $F_\text{wall}\sim h^{-5}$ holds well for all sampled colloid radii. In three-dimensions, the repulsion matches $F_\text{wall}\sim h^{-6}$ for $R = 6$, but experiences a smaller power law for $R = 4$. This indicates that N-MPCD elastic interactions are most accurately resolved for colloid radii $R > 4$. These force measurements demonstrate that N-MPCD accurately simulates long-range quadrupolar deformation in the bulk nematic order and that the colloids dynamically respond to elastic stresses on their surface.

\subsubsection{Force anisotropy}
\label{sctn:aniInt}

While the interactions between quadrupolar colloid-defect complexes and walls are purely repulsive, the long-range interactions between pairs of quadrupolar colloids are more complicated and can alternate between repulsive and attractive depending on relative quadrupole orientation~\cite{musevic2018,chernyshuk2011}. 
To explore this, a 2D colloid is fixed in place (\fig{fig:FDmethods}a) 
while a second mobile colloid is allowed to explore different relative configurations. 
The director is initialised with $\dir=\basis_y$ which forms two $-1/2$ defects beside each colloid and establishes the quadrupole orientations.
Various initial separations and angles are considered (\sctn{sctn:attRepZones} for system and measurement details) and the early time dynamics of mobile colloids are measured. 

The N-MPCD mobile colloid does indeed exhibit regions of both repulsion and attraction. 
The repulsive regions are clearest for pole-to-pole orientations and exist in the far-field limit of small-angle defect-to-defect orientations (\fig{fig:forceDist}c). 
Configurations with intermediate relative angles exhibit attractive interactions. 
Far-field interactions between two quadrupolar colloids separated by a distance $h$ with a relative angle $\theta$ are predicted to have the form 
\begin{align}
    \vec{F}_\text{pair} &\sim K\frac{R^4}{h^5}\cos\left(4\theta\right),
    \label{eq:colloidPair}
\end{align}
in 2D~\cite{dolganov2006,muller2020}.
The sign of the expected interaction force from \eq{eq:colloidPair} show agreement to the simulations, especially in the far-field (\fig{fig:forceDist}c). 

The expectation breaks down at small angles and distances (\fig{fig:forceDist}c). 
The N-MPCD algorithm produces attraction at these sampled points, in contrast with the idealised prediction (\eq{eq:colloidPair}). 
This is partly because the far-field assumptions are less valid but, more importantly, is related to the mechanics of self-assembly: 
The dimer pair quickly self-assembles into a linear chain~\cite{tasinkevych2002,silvestre2004}, causing the colloids to become attractively bound (\fig{fig:ARmethods}c).
Unlike 3D~\cite{skarabot2008}, two-dimensional nematic colloids have a pair of $-1/2$ point defects (\fig{fig:FDmethods}c), which can be freely shared between colloids (\fig{fig:ARmethods}c).
While this section has demonstrated the far-field elastic interactions and a self-assembled 2D chain within N-MPCD, the next section will explore disclination line entanglements between colloidal pairs in 3D. 

\subsection{Entangled defect lines around colloidal dimers}

\begin{figure*}[t]
    \centering
    \includegraphics[width=1\linewidth]{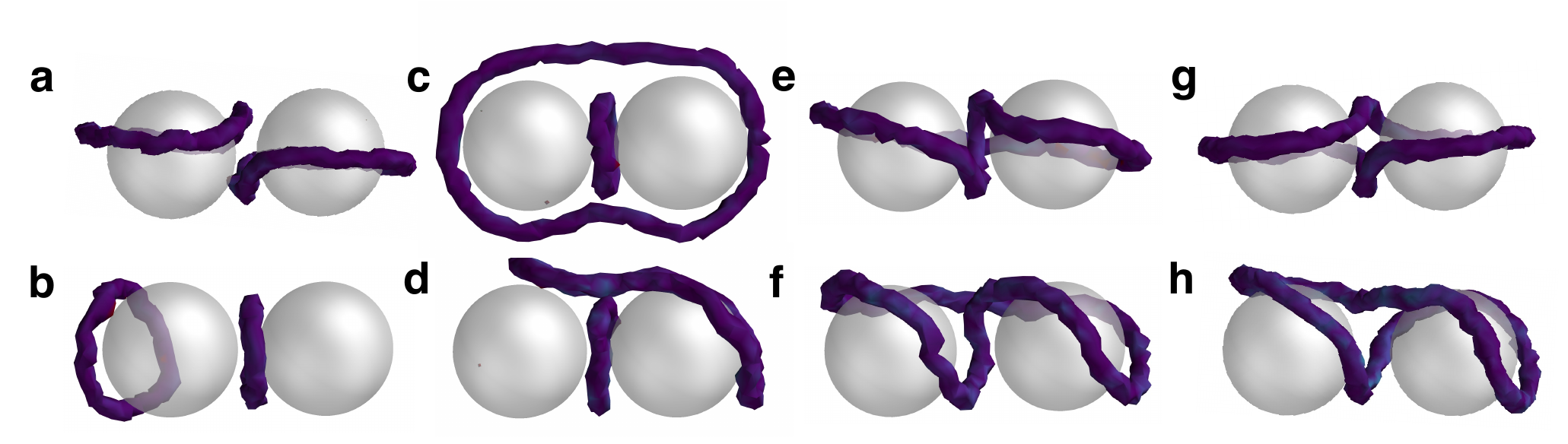}
    \caption{Defect states accompanying colloidal dimers. 
    \textbf{(a)} Saturn rings. 
    \textbf{(b)} Dipolar halos. 
    \textbf{(c)} Figure-of-theta. 
    \textbf{(d)} Tilted-figure-of-theta. 
    \textbf{(e)} Figure-of-omega. 
    \textbf{(f)} Tilted-figure-of-omega. 
    \textbf{(g)} Figure-of-eight. 
    \textbf{(h)} Tilted-figure-of-eight. 
    Disclinations are visualised as in \fig{fig:singleColloid}, confirming each configuration is associated with $-1/2$ disclination loops. 
    }
    \label{fig:entangledstates}
\end{figure*}

Extending into systems with two or more colloids in 3D brings a rich topological interplay between point defects and disclination loops~\cite{alexander2012,alexander2022}, resulting in a range of defect structures including disclination lines that surround multiple colloids~\cite{guzman2003,araki2006}. Entangled states are metastable, and can be induced by a thermal quench~\cite{ravnik2009colloids}, laser manipulation~\cite{ravnik2007}, chiral ordering~\cite{Tkalec2009}, or high colloidal volume fractions~\cite{wood2011}. 
In this study, these states are reached by initialising the bulk fluid from a thermal quench. 
Two mobile colloids are initialised at $\vec{q}_1 = [20,20,13]$ and $\vec{q}_2 = [20,20,27]$, each with homeotropic anchoring in a $L_x=L_y=L_z=40$ domain with periodic boundary conditions on all walls. A warmup phase is applied ($T_W = 200$) where nematic order forms and no data is collected. Simulations are then run for the duration $T_S= 1000$.
Loops are identified via the disclination density tensor (\sctn{sctn:defects}~\cite{schimming2022}).
Eight disclination states are observed from the N-MPCD simulations with either one or two $-1/2$ disclination loops (\fig{fig:entangledstates}). 
Of these, the two states that are not considered entangled are extensions of the single colloid case, with either two Saturn rings or two dipolar halos that assemble into a chain. The others are entangled with at least one loop ($n\geq1$) that wraps around the colloidal dimers. These states derive their names from the shape of their disclinations.
In the case of the figure-of-theta (\fig{fig:entangledstates}c), two loops exist ($n=2$): one large ring that encircles both colloids, and another smaller ring positioned between them. The figure-of-omega (\fig{fig:entangledstates}e) and figure-of-eight (\fig{fig:entangledstates}g) are single loop entanglements ($n=1$). Each of these states have been well-documented in experiments and simulations~\cite{ravnik2007,ravnik2009colloids,tkalec2013}.

Additionally, the N-MPCD algorithm reveals the existence of tilted analogues of the figure-of-theta (\fig{fig:entangledstates}d), figure-of-omega (\fig{fig:entangledstates}f) and figure-of-eight ((\fig{fig:entangledstates}h). 
These are tilted with respect to the axis the colloids reside in. 
While rare, these tilted entangled dimer states emerge when the director field does not form a uniform alignment axis away from the colloids (see \fig{fig:farfield}).
This generates modulated order that cannot relax to the ground state. In these simulations, the combination of colloids, periodic boundary conditions and quenched disorder are able to trap these tilted entangled states.  

\begin{figure}[tb]
    \centering
    \includegraphics[width=\linewidth]{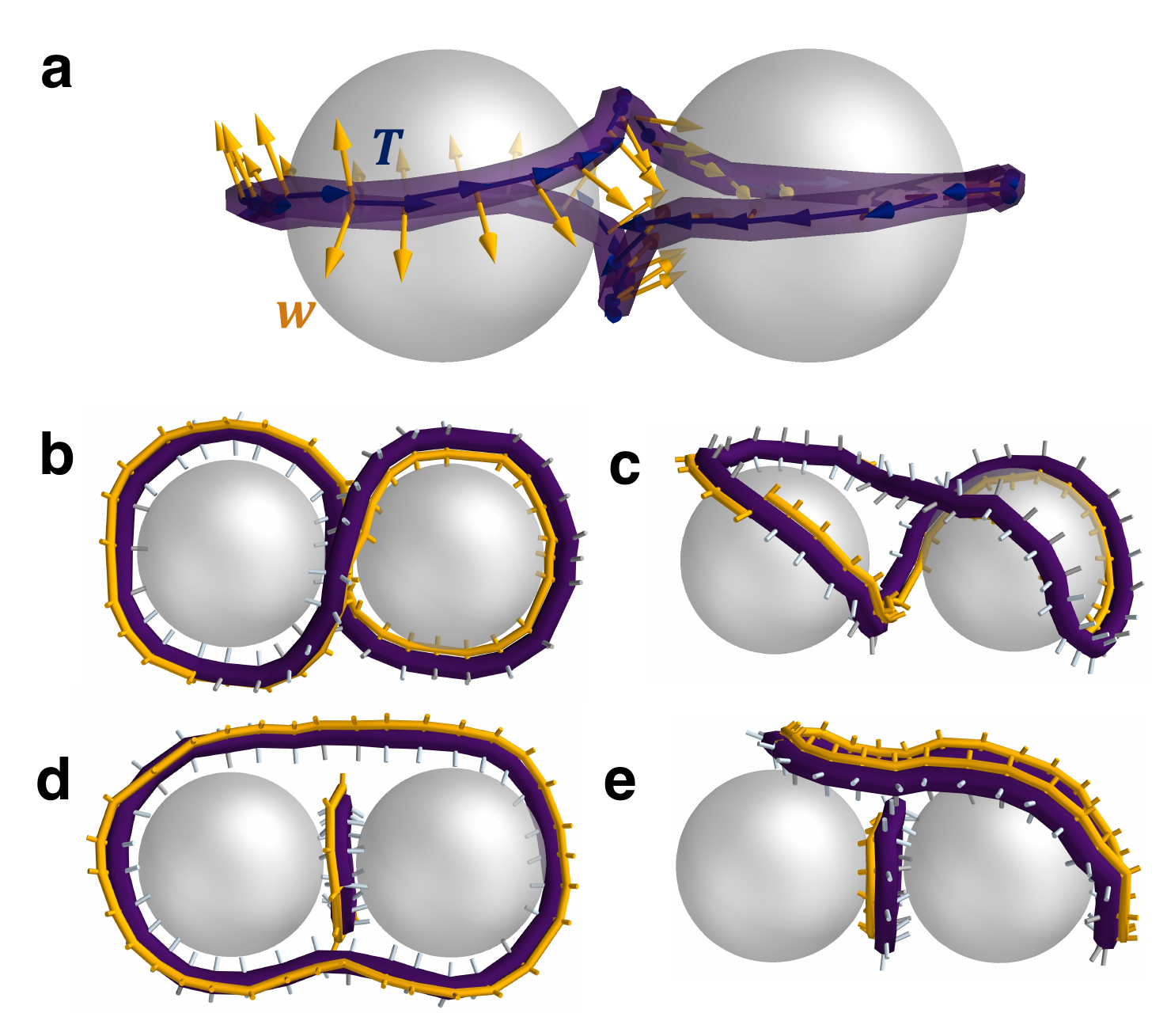}
    \caption{Minus-half disclination loops as ribbons with self-linking numbers $\Sl$. 
    \textbf{(a)} Figure-of-eight presented as 
    ribbons constructed through identifying the disclination line tangent $\tanVecdG$ (navy blue arrows) and local framing vector $\framing$ (orange arrows). 
    Disclination loops are as in \fig{fig:singleColloid}.
    \textbf{(b)} Figure-of-eight from \textbf{(b)} as viewed from second perspective. 
    \textbf{(c)} Tilted-figure-of-eight. 
    \textbf{(d)} Figure-of-theta. 
    \textbf{(e)} Tilted-figure-of-theta. 
    In each of \textbf{(b)}-\textbf{(e)}, the orange framing curve smoothly connects the -1/2 wedge orientations denoted by orange cylinders. 
    Silver cylinders illustrate the other two radially outward pointing orientations. 
    In \textbf{(b)}-\textbf{(c)}, the end points do not meet the starting points of the orange framing curve, which indicates a net rotation and implies Sl$\neq0$. 
    On the other hand, the orange framing curve is continuous and Sl$=0$ in \textbf{(d)}-\textbf{(e)}. 
    }
    \label{fig:ribbonColloids}
\end{figure}

With the disclination states identified, we next characterise their topological and geometric properties. To obtain these, the framework by \v{C}opar and \v{Z}umer is followed \cite{copar2011n1,copar2011n2}.
Since colloidal anchoring enforces a geometric constraint for the local director to lie in a plane perpendicular to $\tanVecdG$ ($\cos\beta=-1$, in this case), the disclination loop can be assigned a framing vector $\framing$ that is everywhere perpendicular to the tangent (\fig{fig:ribbonColloids}a; see \sctn{sctn:ribbon}). 
A convenient choice of $\framing$ is one of the three radially pointing director orientations of the $-1/2$ disclination (\fig{fig:ribbonColloids}a). 
The framing vector allows the topological properties of the $-1/2$ disclination loop to be found via the self-linking number $\Sl$, which counts the number of times the framing turns around
the tangent on traversing the loop.
The self-linking number can be calculated from geometric properties of the disclination
through the C\v{a}lug\v{a}reanu-White-Fuller theorem 
\begin{align}
    \label{eq:sl}
    \Sl &= \Wr+\Tw, 
\end{align}
where $\Wr$ is the writhe and $\Tw$ is the twist (\sctn{sctn:self-linking}). 
Due to the three-fold symmetry of $-1/2$ disclinations, $\Sl$ takes fractional, third-integer values. The self-linking number is related to the topological classification of $-1/2$ disclination loops through \cite{copar2013}
\begin{align}
    \label{eq:mappingtoSl}
    \nu = 3\Sl + 2 \quad\text{(mod 4)},
\end{align}
where $\nu$ is the topological index of a disclination loop \cite{janich1987}. Index values of $\nu=0$ correspond to unlinked and charge neutral ($p=$even), $\nu=2$ unlinked and charged ($p=$odd) and $\nu=1,3$ are linked loops. 
In this way, the relationship between $\Wr$, $\Tw, \Sl$ and point charge $p$ can be understood for the N-MPCD $-1/2$ disclination states in \fig{fig:entangledstates}.
\begin{table*}
\small
  \caption{\ Classification of the eight identified nematic disclination states in terms of topological and geometric information. Disclination properties include the writhe ($\Wr$) and twist ($\Tw$), which combine to give the topologically-protected self-linking number $\Sl=\Wr+\Tw$. 
    Topological point charge $\pointcharge$ associated with colloidal dimers combine to give a trivial nematic texture (even), allowing even contributions from each $n=1$ or two odd contributions for states with $n=2$ loops. The properties are calculated directly from the frames shown in \fig{fig:entangledstates}.
    }
  \label{tab:classification}
  \begin{tabular*}{\textwidth}{@{\extracolsep{\fill}}lllllll}
    \hline
     & $n$ & Wr & Tw & Sl & $\pointcharge$\\
    \hline
    Saturn rings & 2 & 0.014 \vline \  0.001 & 0.029 \vline \ 0.053 & 0.044 \vline \ 0.054 & odd \vline \ odd\\
    Dipolar halos & 2 & 0.009 \vline \ 0.003 & 0.048 \vline \ 0.003 & 0.057 \vline \ 0.005 & odd \vline \ odd\\
    Figure-of-Theta & 2& 0.035 \vline \ 0.002 & 0.025 \vline \ -0.011 & 0.060 \vline \ -0.008 & odd \vline \ odd\\
    Figure-of-Eight & 1 & 0.699 & -0.027 & 0.673 & even\\
    Figure-of-Omega & 1 & 0.652 & 0.009 & 0.661 & even\\
    Tilted-Figure-of-Theta & 2 & 0.005 \vline \ 0.005 & -0.008 \vline \ 0.013 & -0.003 \vline \ 0.018 & odd \vline \ odd\\
    Tilted-Figure-of-Eight & 1 & -0.705 & 0.051 & -0.654 & even\\
    Tilted-Figure-of-Omega & 1 & 0.618 & 0.065 & 0.683 & even\\
    \hline
  \end{tabular*}
\end{table*}

First, the properties for the entangled single loop ($n = 1$) states are examined. For the figure-of-eight, figure-of-omega and their tilted analogues, the self-linking number is found to be $\Sl \approx ±2/3$ (\tbl{tab:classification}). Additionally, the $\Sl \approx ±2/3$ can be visualised for the two figure-of-eight states by tracking the orientation of the $-1/2$ profile (\fig{fig:ribbonColloids}b,c). In choosing a reference and tracking the profile rotations along the loop (orange ribbon curve), the orientation is rotated by $\pm2\pi/3$ over the entire contour of the loop. For each $n=1$ state, the $\Sl$ is composed entirely from writhe, while the twist remains essentially zero in each state (\tbl{tab:classification}). Self-linkings composed entirely of writhe were previously observed for the figure-of-eight and figure-of-omega~\cite{copar2011n1}, since the strong radial constraint on the disclination profile penalises twisting of the orientation. We show the same writhe/twist balance also hold when the disclinations are in tilted conformations. The $\pm$ sign on the Sl relates only to the chirality of the conformation and does not influence the topological classification of the loop. Indeed, mapping to the disclination loop index reveals that all four states are topologically trivial $\nu=0$ (uncharged with $p =$ even). The $n = 1$ disclination line balances the two point charges provided by the colloids by forming a state with net writhe $\Wr$. 

Next, the disclination states with $n=2$ are examined. Each state has a self-linking of $\Sl \approx 0$ (\tbl{tab:classification}). This is the case for the individual rings (Saturn rings and dipolar halos) and the entangled figure-of-theta structures, each presenting $\Wr \approx 0$ and $\Tw\approx0$. We visualise the ribbon (orange curve) for the two figure-of-theta states in \fig{fig:ribbonColloids}d,e, which confirm the calculated properties. The reference orientation smoothly connects to the final orientation, with no local (or global) twisting or coiling over the circuit. The $\Sl = 0$ properties finds that each loop carries a hedgehog charge $p =$ odd ($\nu=2$), balancing the global charge neutrality between the two loops (modulo 2). These results show that each $n = 2$ state is topologically equivalent with identical geometric decomposition into $\Wr = 0$ and $\Tw = 0$. Therefore, the tilted states are simply smooth transformations of their non-tilted counterparts. 

\subsection{Entanglement kinetics}

With each of the disclination states identified and characterised, we study the relaxation pathways that lead to the formation of these states. 
As already demonstrated for a single colloid (\fig{fig:singleColloid}d), disclinations contour lengths generally decrease as the system relaxes from the thermal quench.
For dimers, the temporal evolution of the disclination contour lengths eventually leads to the long-time configurations from \fig{fig:entangledstates}. 
Since N-MPCD simulates fluctuating nematohydrodynamics, the simulations stochastically sample states as they relax towards accessible lower free energy configurations. 

\begin{figure*}[tb]
    \centering
    \includegraphics[width=0.75\linewidth]{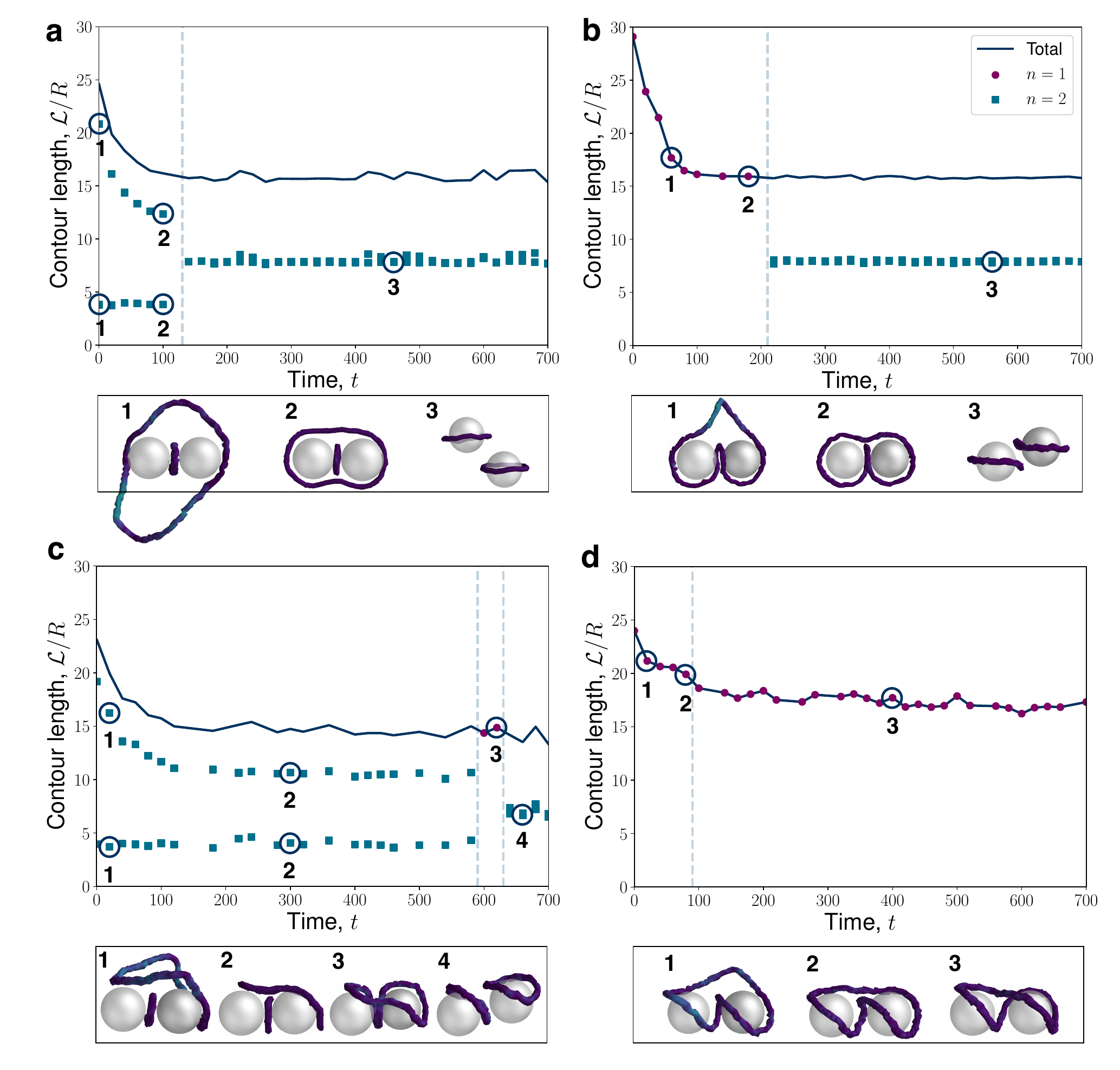}
    \caption{
    Relaxation pathways for dimer-associated disclination loops following a thermal quench, as measured via the disclination contour lengths $\mathcal{L}$ scaled by colloid radus $R=6$. 
    \textbf{(a)} Figure-of-theta transitioning to Saturn rings. 
    \textbf{(b)} Figure-of-omega transitioning to Saturn rings. 
    \textbf{(c)} Tilted figure-of-theta transitioning to dipolar-halo pair, passing briefly through a tilted figure-of-omega state. 
    \textbf{(d)} Tilted figure-of-omega transitioning to tilted figure-of-eight. 
    Topological transitions via loop reconnection or splitting events are indicated by vertical dashed lines. 
    In each panel, the time $t=0$ corresponds to the first recorded timestep for which the largest disclination loop is entirely contained within the periodic system. 
    Circles denote a single loop ($n=1$) and two loops ($n=2$) are shown as square markers, while the total contour length is shown as the navy blue line. 
    Example snapshots are shown directly below each panel.
    }
    \label{fig:contourLength}
\end{figure*}

Four instances of the stochastic relaxation of the entangled dimers are shown in \fig{fig:contourLength}. 
An example of the relaxation passing through a figure-of-theta is shown in \fig{fig:contourLength}a. 
At early times (\fig{fig:contourLength}a.1), a small loop exists sandwiched between the colloids with a contour length $\mathcal{L}$ comparable-to-but-less-than the circumference of the colloids. 
Simultaneously, a large disclination loop rapidly collapses around the colloids, forming the figure-of-theta state (\fig{fig:contourLength}a.2). 
The number of loops is $n=2$ throughout. 
In N-MPCD, the figure-of-theta is only sampled transiently, passing rapidly through loop-reconnections to form two Saturn ring colloids (\fig{fig:contourLength}a.3). 
Despite sharp transitions in the individual loop lengths (\fig{fig:contourLength}a), the total contour length has a negligible change between the two states --- 
with two equal-sized Saturn rings that sum to the total disclination length of the two figure-of-theta loops.

Another kinetic trajectory observed in N-MPCD is a single ($n=1$) quenched disclination loop (\fig{fig:contourLength}b.1) that collapses to form a figure-of-omega state (\fig{fig:contourLength}b.2). 
The figure-of-omega entangled state is found to be metastable with a constant contour length for $t\approx100$, after which time the entangled loop transitions to two Saturn rings (\fig{fig:contourLength}b.3). 
Unlike the transition from the figure-of-theta state in \fig{fig:contourLength}a, the transition from the figure-of-omega state involves a topological conversion from $\Sl\approx2/3$ to two rings with $\Sl\approx0$ (\tbl{tab:classification}). Equivalently, this corresponds to a transition from a single uncharged loop, to two charged loops.

The tilted entanglements can show somewhat different trajectories because of their non-uniform global director alignment (\fig{fig:contourLength}c). 
The tilted state arises because the disclination collapses at an off-set to the colloidal axis (\fig{fig:contourLength}c.1), passing into the tilted-figure-of-theta (\fig{fig:contourLength}c.2). 
The tilted figure-of-theta state endures for an extended time ($100\leq t\leq 200$) with minimal changes to the conformation, until a segment of the disclination line reconnects into a fleetingly brief tilted-figure-of-omega state (\fig{fig:contourLength}c.3). Finally, the disclination divides into two dipolar halos with orientations tilted with respect to each other (\fig{fig:contourLength}c.4). 

An $n=1$ tilted relaxation trajectory can also occur, starting with a larger loop (\fig{fig:contourLength}d.1) that encloses the colloid pair to form the tilted-figure-of-omega state (\fig{fig:contourLength}d.2). 
As in \fig{fig:contourLength}c, this tilted-figure-of-omega state is short lived and, in this case, transitions to the tilted-figure-of-eight without transitioning through $n=2$ (\fig{fig:contourLength}d.3). 
Interestingly, the tilted-figure-of-eight is observed to be the most stable of any of the entangled states observed in N-MPCD simulations, remaining in the same configuration for the entire simulation, with minimal variation in contour length. This parallels experimental observations~\cite{jampani2011}, albeit for different states and surrounding order, where chirality or modulated order can offer protection from reaching the global free energy minimum \cite{tkalec2011}. In addition, figure-of-eights have been associated with the
greatest stability of all entangled structures \cite{ravnik2007}.
Despite the intrinsic stochasticity of the numerical approach, the tilted-figure-of-eight was not observed to relax into states with $n=2$ rings.

\begin{figure}[tb]
    \centering
    \includegraphics[width=1\linewidth]{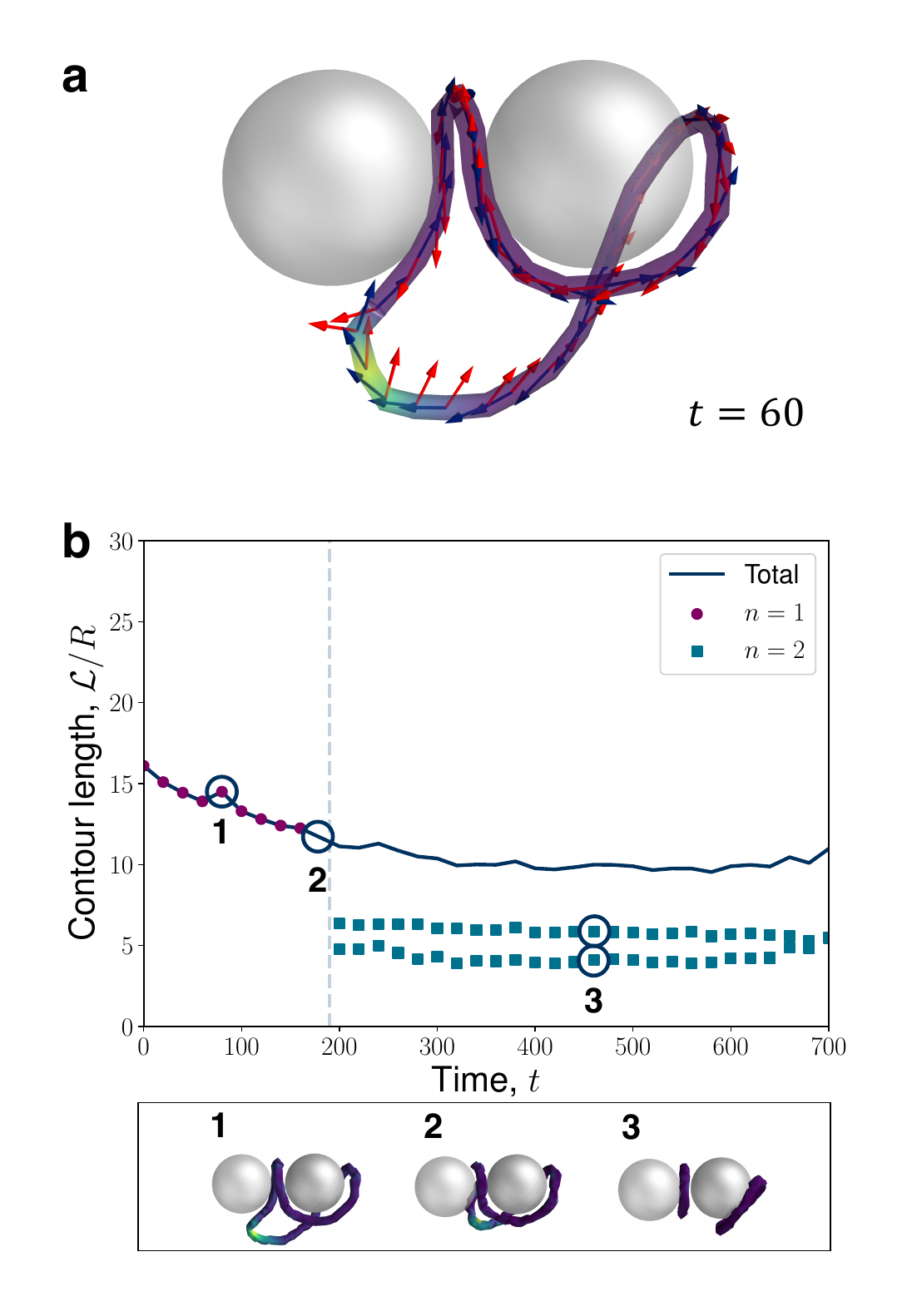}
    \caption{Quenched disorder can sample out-of-equilibrium disclination configurations. 
    \textbf{(a)} Early-time snapshot of a charge-neutral disclination with a localised $+1/2$ profile (yellow segment), visualised as in \fig{fig:singleColloid}. 
    \textbf{(b)} Relaxation trajectory of the normalised contour length $\mathcal{L}/R$ with time $t$. 
    Markers, lines and colouring the same as \fig{fig:contourLength}. Example snapshots are shown directly below the panel.
    }
    \label{fig:nonminushalf}
\end{figure}

\begin{figure*}[tb]
    \centering
    \includegraphics[width=0.8\linewidth]{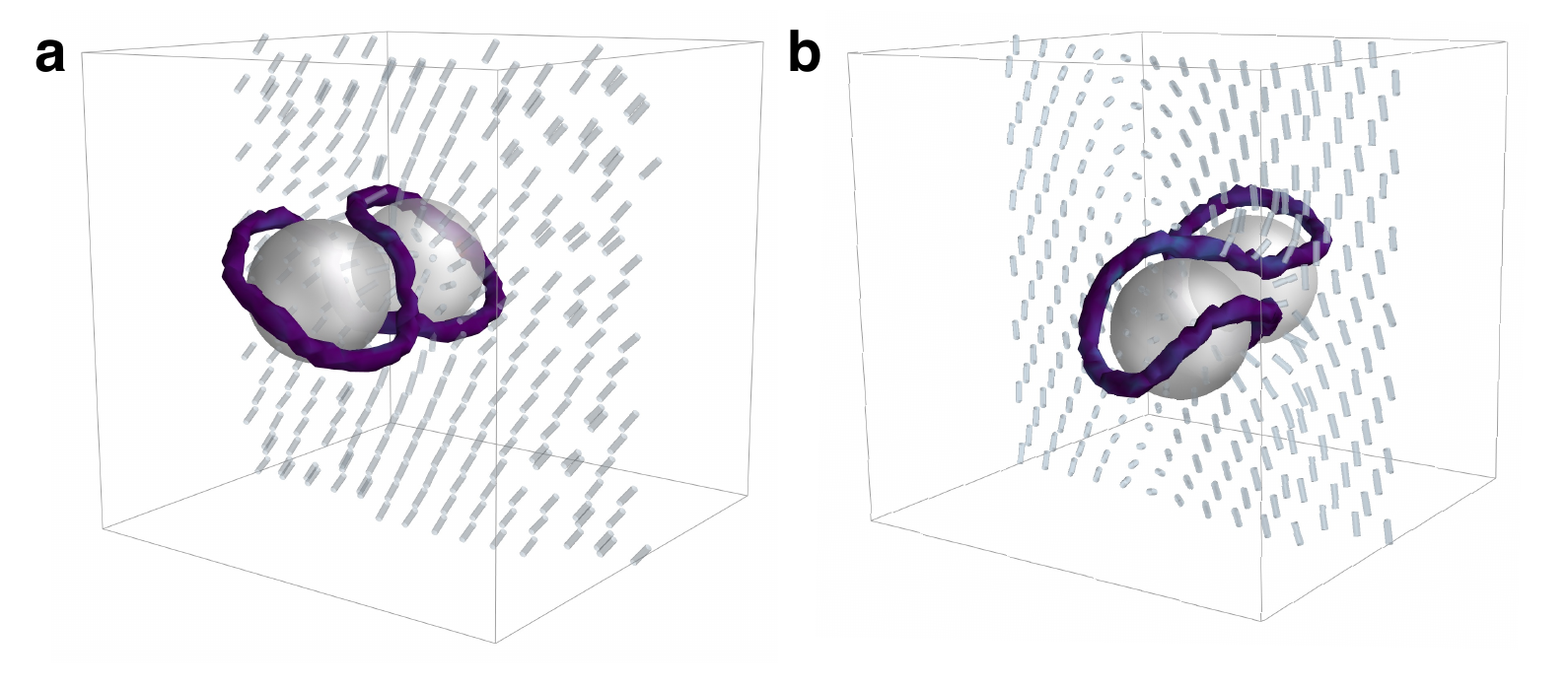}
    \caption{Director field slices around standard and tilted figure-of-eight colloidal-dimer states. 
    \textbf{(a)} Standard figure-of-eight structures are associated with a uniaxial far-field director field. \textbf{(b)} Tilted-figure-of-eight entanglements have a director field that modulates away from the dimer. 
    The director is shown in grey. 
    Disclinations are visualised as in \fig{fig:singleColloid}.
    }
    \label{fig:farfield}
\end{figure*}

Infrequently, less conventional entangled-dimer relaxation dynamics are revealed by N-MPCD, such as the situation shown in \fig{fig:nonminushalf}. Similar to the tilted structures, this trajectory eventually relaxes into a modulated global director field (\fig{fig:farfield}). 
However, at early times, an unexpected entangled state emerges in which the disclination loop has a localised segment with a $+1/2$ wedge profile  (\fig{fig:nonminushalf}a). 
The $+1/2$ profile is smoothly connected by fleeting twist to a majority $-1/2$ loop.
This wedge-twist state necessarily contains $\pointcharge=$even to balance the charge of the dimers.
Generally, such $+1/2$ wedge profiles are discouraged since the global director alignment cannot coexist with the low-symmetry of the $+1/2$ wedge and out-of-plane twist is penalised by the radial colloidal anchoring. 
In this case, the penalty against twist is resolved by a rapid reorientation of the rotation vector $\rotVecdG$, which rotates by $\pi$ relative to the global basis over a small disclination segment (\fig{fig:nonminushalf}a). 
The $+1/2$ segment of the disclination gradually approaches the colloids (\fig{fig:nonminushalf}b.2) until it combines with a $-1/2$ profile, facilitating a topological transition from a single loop ($n=1$) to a state with a pair of $-1/2$ dipolar halos (\fig{fig:nonminushalf}b.3). 

\section{Conclusions}

This work has utilised Nematic Multi-Particle Collision Dynamics (N-MPCD) to simulate nematic colloids as mobile surfaces that can resolve stresses at the interfaces. 
In three-dimensions, N-MPCD reproduces the experimentally observed and theoretically predicted colloid-disclination complexes for solitary colloids. 
These include \textit{(i)} Boojums with handle-shaped semi-loops, \textit{(ii)} Saturn rings and \textit{(iii)} dipolar halos. 
Furthermore, N-MPCD mediates elastic interactions between colloidal inclusions. 
The elastic forces in N-MPCD are seen to decay with the expected power-laws in two- and three-dimensions. Likewise, the anisotropy of quadrupoles interacting in the far-field-limit has been demonstrated for colloids and their accompanying pairs of free point defects in 2D. 
If the colloids are too near to each other, the far-field approximation breaks down and N-MPCD predicts that dimer structures are formed through shared point defects. 
For nearby colloidal dimers subjected to a 3D thermal quench,  N-MPCD reproduces expected defect structures, including disclination loops that entangle both colloids. 
In addition to the expected defect structures, previously unobserved analogous tilted entanglements are revealed by N-MPCD in systems with periodic boundary conditions. 
In these tilted states, the far-field directors are not uniform compared to the previously obsserved states. 

Despite being a noisily fluctuating algorithm, N-MPCD not only respects topological constraints but also resolves details of defect topology and disclination structure, such as self-linking numbers or localised wedge/twist profiles. 
Furthermore, as a linearised nematohydrodynamic approach, N-MPCD simulates the entanglement kinetics. 
This allows the algorithm to 
explore relaxation from a quench --- revealing that topological point charge is not evenly distributed around the loop, but instead carried by segments of the disclination loop closest to the colloidal surface. 
This illustrates that N-MPCD is ideal for accessing and exploring metastable states, owing to the intrinsic thermal noise and dynamics beyond overdamped free-energy steepest descent. 
In particular, the simulations produced an early-time charge-neutral disclination state that does not conform to an entirely -1/2 disclination loop. 

This study demonstrates that the N-MPCD algorithm is well-suited for studies on topological kinetics, field-driven assembly and colloidal self-assembly. 
The versatility of combining complex embedded~\cite{hijar2020} or confining geometries~\cite{Wamsler2024}, fluctuating nematohydrodynamic flows and out-of-equilibrium dynamics~\cite{kozhukhov2022} makes N-MPCD highly suitable coarse-grained approach for studying dynamics of topological phenomena. Further work could apply the N-MPCD algorithm to study the interactions and defect structures surrounding nematic colloids in active nematic systems, or topological features of the percolated -1/2 disclination loops in colloid nematic gels \cite{wood2011}. 
The control over complex surfaces could be used to explore colloids in complex geometries, including the possibility of kinetics and fluctuations in non-trivial knotted fields~\cite{Machon2019}.
This work contributes to a numerical approach to study the relationship between topology and rheological properties.

\section*{Conflicts of interest}
There are no conflicts to declare.

\section*{Acknowledgements}
This research has received funding from the European Research Council under the European Union’s Horizon 2020 research and innovation programme (Grant
Agreement Nos. 851196). 
For the purpose of open access, the author has applied a Creative Commons Attribution (CC BY) licence to any Author Accepted Manuscript version arising from this submission.

\section{Appendix}
\subsection{Kleman–de Gennes extrapolation length}
\label{appendix:klemen-deGennes}

\begin{figure}[t]
    \centering
    \includegraphics[width=0.8\linewidth]{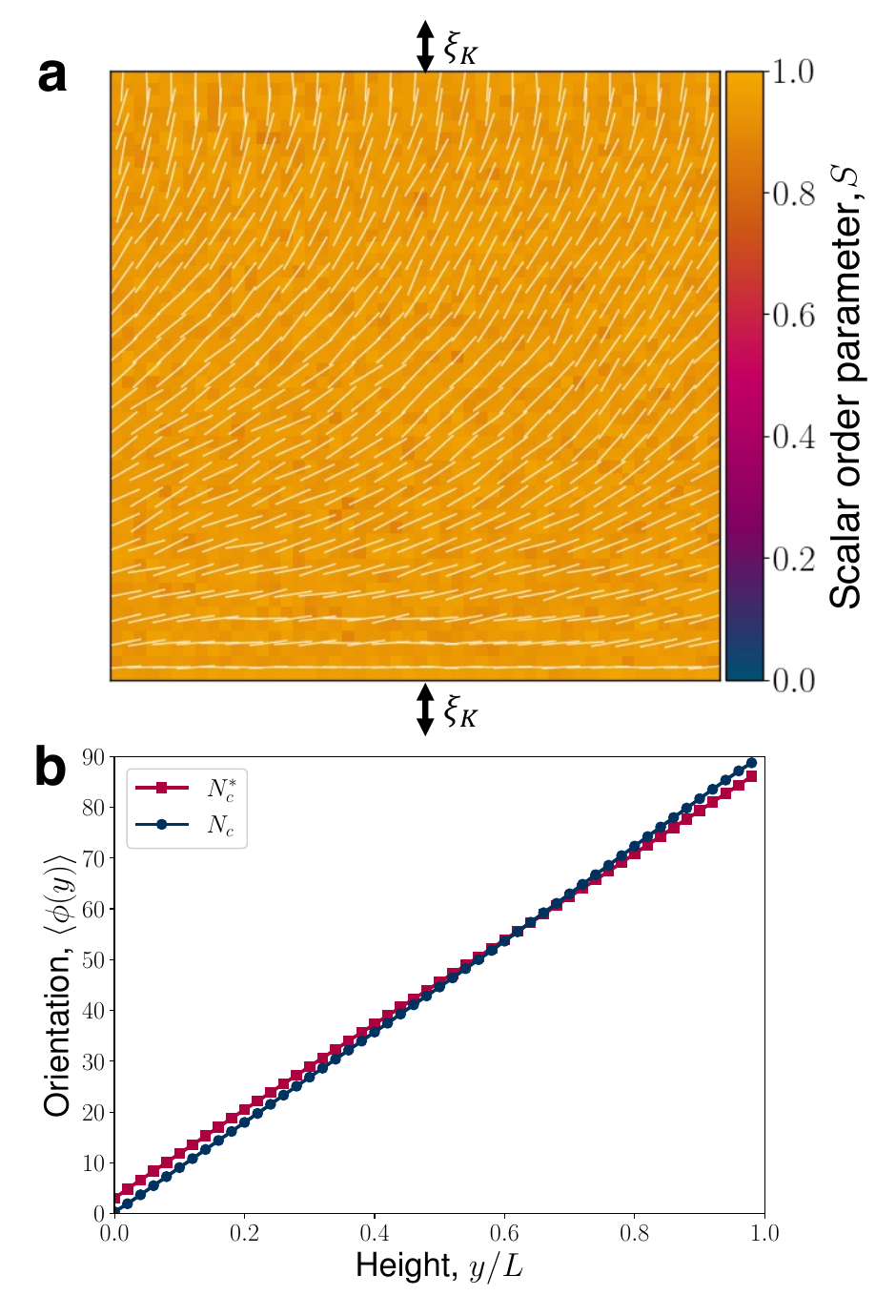}
    \caption{Extrapolation length. 
    \textbf{(a)} A hybrid-aligned nematic cell of size $L=50$ with planar anchoring on the bottom surface and homeotropic on the top, and periodic boundaries on the sides. 
    The director field is shown as the white line field. 
    For finite strength anchoring, the anchoring starts an effective distance $\extrapolationlen$ behind each surface. 
    \textbf{(b)} Director orientation $\langle \phi(y) \rangle $ with vertical height $y$, averaged over horizontal position $x$, time $t$, and $N=20$ simulation runs. 
    Red squares present orientation when only  the $N^*\leq N_c$ particles that directly violate \eq{eq:surfviolation} have the anchoring conditions applied.  
    Dark blue circles are for for transformations applied to all particles within a cell $N_c$.
    }
    \label{fig:extrapolation}
\end{figure}  

The extrapolation length is a length scale that measures the competing influence of elasticity against anchoring strength $\extrapolationlen=K/W$, where $K$ is the elastic constant under the one constant approximation, and $W$ is the anchoring strength \cite{de1993physics}. 
To compare the influence of anchoring applied only to proportion of particles $N^*_c/N_c$ against the stronger anchoring method described in \sctn{sctn:BCs}, applied to all $N_c$ particles in cells that intersect the surface, the extrapolation length is measured in a two-dimensional hybrid-aligned-nematic cell with homeotropic anchoring on the top boundary $y=L$ and planar anchoring on the bottom boundary $y=0$ (\fig{fig:extrapolation}a). 
In the $\basis_x$ direction, periodic boundary conditions are applied. 
The system size is $L_x=L_y=L=50$, with 20 simulation runs of $10^4$ time steps, outputted every $100$ timesteps to establish an average for the orientation angle $\phi(y)$ (defined from the positive $\basis_x$ axis) over all runs and timesteps. 

Assuming that the anchoring condition begins a small distance $\extrapolationlen$ beyond each boundary, so that the nematic orientation 
$\phi(-\extrapolationlen)=0^\circ$ and $\phi(L+\extrapolationlen)=90^\circ$, the extrapolation length can be found from
\begin{align}
    \extrapolationlen = \frac{1}{2}\left(\frac{90}{m_{g}}-L\right),
\end{align} 
where $m_g$ is the gradient of the linear fit in degrees per unit length.
This gives $\extrapolationlen=0.147\pm0.0002$ for the $N_c$ case, and $\extrapolationlen=1.700\pm0.001$ for the $N^*_c$ case (\fig{fig:extrapolation}b).
For a colloid with radius $R$, the strength of the anchoring is given by the dimensionless ratio of the surface free energy $WR^2$ cost against elastic energy $KR$, which produces a reduced colloid size, $RW/K=R/\extrapolationlen$. 
All simulations use the strong anchoring method ($N_c$ case). Three-dimensional colloids in this paper have a radius of $R=6$ in simulation units, giving $R/\extrapolationlen=40.8$. In two-dimensions, $R=10$ gives $R/\extrapolationlen=68.0$.

\subsection{Anchoring torque}
\label{SI:anchoringtorque}

The particle orientation transformations described in \sctn{section:mobilecolloids} are implemented as hard anchoring conditions that align MPCD particle $i$. 
The initial orientation of the particle prior to colliding with the surface is $\ori_i$. 
The change in nematogen orientation due to the collision is $\delta \ori_i $. 
This orientational change must be converted into a  force on the colloid. 
We infer the torque $\vec{\torque}_i^{\mathrm{anch}} = \gamma_R \ori_i \times \frac{\delta \ori_i}{\delta t}$ as the rotation through the fluid with rotational friction coefficient $\gamma_R$.
The final particle orientation, post anchoring, can be written in terms of the scalar multipliers as $\ori^{\mathrm{final}} = (\mathcal{M}_{\surfnormal_b} \surfnormal_b + \mathcal{M}_{\surftangent_b} \surftangent_b)\left( \mathcal{M}_{\surfnormal_b}^2+\mathcal{M}_{\surftangent_b}^2 \right)^{-1/2}$. 
Taking the cross product gives \eq{eq:anchoringtorque}. 
One caveat to inferring the torque in this manner, is that the periodicity of $(\ori_i\cdot \surfnormal_b)(\ori_i\times \surfnormal_b)$ only infers the correct torque magnitude for angles $-\pi/4\leq\alpha\leq\pi/4$. In the N-MPCD simulations presented here, reorientations greater than $\pi/4$ are rare.

\subsection{Torque to force}
\label{SI:torquetoforce}

The elastic force exerted on a colloid (mobile boundary), due to the anchoring transformation of a single N-MPCD particle, is determined by the torque on a virtual N-MPCD particle (\sctn{section:mobilecolloids}). This torque conserves angular impulse (\eq{eq:anchtorquebalance}). 
Since the torque is a pseudovector, converting a torque into a force is not generally possible --- there can be colinear contributions between the force $\vec{F}$ and radial vector $\vec{r}$ that return the same value of torque $\torque$. The non-unique nature of the force is demonstrated by the identity
\begin{align}
    \label{eq:identity}
    \vec{F} &= \frac{\vec{r}\left(\vec{r}\cdot \vec{F}\right) - \vec{r}\times\torque}{r^2}.
\end{align}
Since the anchoring torque is a purely rotational effect, we assume that the colinear contribution of the force is zero ($\vec{r}\cdot\vec{F}=0)$. Under this assumption of orthogonality between $\mathbf{r},\vec{F},\torque$, the first term in the numerator of \eq{eq:identity} vanishes and so force can be inferred from torque. 
In \eq{eq:torque2force}, the force is $\force=\force_{b,i}$, the torque is $\torque=\torque_{b,i}$ and the radial vector is $\vec{r}=(\ell_u/2)\ori_{b,i}$ which corresponds to half of the nematogen rod length.

\begin{figure*}[tb]
    \centering
    \includegraphics[width=1\linewidth]{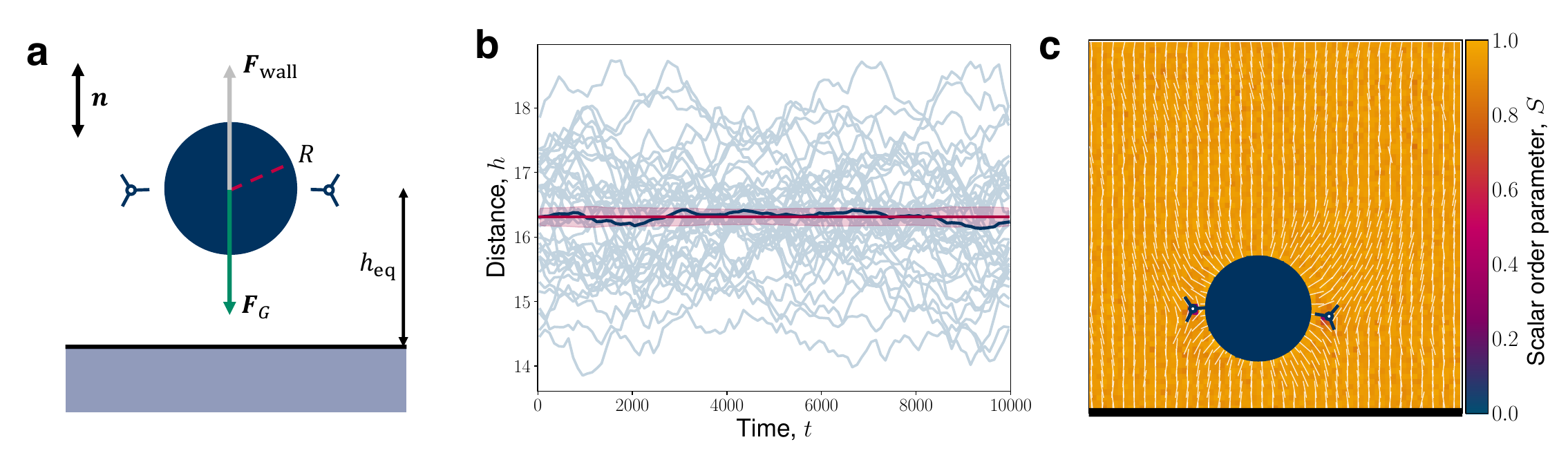}
    \caption{Measurements for determining the force-distance relation (\fig{fig:forceDist}). 
    \textbf{(a)} Schematic of colloid (navy circle) with radius $R$, initialised within the far-field director $\dir$, forming equatorial -1/2 defects (blue trefoil symbols). 
    The colloid experiences an elastic repulsion force due to the wall $\vec{F}_{\text{wall}}$, which is probed by a gravitational-like external force $\vec{F}_{\text{G}}=-M_c G\surfnormal_\mathrm{wall}$.
    The distance $h$ between the colloid’s centre of mass and the wall is $h_{\mathrm{eq}}$ when $\vec{F}_{\mathrm{G}}=\langle \vec{F}_{\text{wall}}\rangle$.
    \textbf{(b)} An example of $30$ colloidal trajectories (grey) for a two-dimensional colloid with radius $R=10$. 
    The time-dependent ensemble mean is shown by the dark blue curve, which fluctuates about the time-averaged mean (red horizontal line). 
    The red shading is the standard error about the time-averaged mean. 
    \textbf{(c)} An example snapshot of a colloid interacting with the homeotropic anchored wall (black horizontal line). The white line field is the director and background colouring is the scalar order parameter $S$.}
    \label{fig:FDmethods}
\end{figure*}
\subsection{Methods for colloid-wall repulsion}
\label{sctn:wallSI}

Systems in two-dimensions have $L_x=L_y=L$ with periodic boundary conditions in $\basis_x$, and solid walls in $\basis_y$. Both upper and lower solid walls have no-slip boundary conditions, but only the lower boundary has anchoring conditions applied, with strong homeotropic alignment (\sctn{appendix:klemen-deGennes}). Four colloid radii are sampled $R\in[6,8,10,12]$ with system sizes $L\in[50,55,60,80]$ respectively, to adjust for system-size effects. 
The director field is initialised from $\dir(t=0)=\basis_y$, which produces a pair of near-surface point defects with $-1/2$ charge in 2D. 
A simulation warmup time of $T_W = 1000$ is applied, during which the colloid is held fixed and the director relaxes to the equilibrium configuration.
Simulations are then performed for $T_S = 30000$, with the colloid mobile and responsive to the nematic environment. A total of 40 independent simulation runs are performed for each $R$.
In three-dimensions, two colloid radii are used. The first is $R = 4$ with $L_x = L_y = L_z = 30$, and the second is $R = 6$ with
$L_x = L_y = 30, L_z = 35$. Simulations have periodic boundary conditions in $\basis_x$ and $\basis_y$, and impermeable no-slip walls in $\basis_z$. Similar to two-dimensions, only the lower plate has homeotropic anchoring conditions. The director field is intialised along $\dir(t=0)=\basis_z$, leading to a quadrupolar Saturn ring. The simulations run for $T_S = 4000$ following a warmup period of $T_W = 1000$, where the colloid is held
static. Statistics are generated from $30$ independent measurements for each $R$.

The decaying power-law nature of the elastic forces are determined by measuring the interaction forces of a nematic colloid with a centre of mass distance $h$ away from a homeotropic anchored wall. In the proximity of the wall, the colloid experiences a strong elastic repulsive force, $\vec{F}_{\mathrm{wall}}$, that decays with distance. This can be represented as a quadrupole-quadrupole elastic interaction between the colloid and its mirror image on the other side of the wall (\eq{eq:colloidWall}).
In addition, the motion of the colloid through the fluid experiences a drag force due to viscosity $\vec{F}_{\mathrm{drag}}$ and a fluctuating force $\vec{F}_{\mathrm{fluc}}$ that enters due to the stochasticity of the collision operators.
To measure $\vec{F}_{\mathrm{wall}}$, we apply an external gravitational-like body force to the colloid
\begin{align}
    \vec{F}_{\mathrm{G}}=M_c\vec{G},
    \label{eq:bodyforce}
\end{align}
where $M_c = \frac{4}{3}\pi R^3 \langle N_c \rangle$ is the mass of the colloid in 3D or $M_c = \pi R^2 \langle N_c \rangle$ in 2D. The constant acceleration $\vec{G}=-G\surfnormal_\mathrm{wall}$ is directed towards the homeotropic wall with surface normal $\surfnormal_\mathrm{wall}$ and magnitude $G$. The applied body force (\eq{eq:bodyforce}) probes the elastic force by introducing an equilibrium distance $h_\mathrm{eq}$ at which $\vec{F}_{\mathrm{wall}}+\vec{F}_{\mathrm{G}}=0$ (\fig{fig:FDmethods}a). 
When the elastic and applied forces balance, the colloid only fluctuates about $h_\mathrm{eq}$ (grey trajectories in \fig{fig:FDmethods}b). Therefore, the fluctuating and drag forces can be neglected in the force measurements provided there is statistical certainty on $h_\mathrm{eq}$. For this reason, the simulations are iteratively re-initialised from new start positions $h \approx h_\mathrm{eq}$ (example director configuration in \fig{fig:FDmethods}c), so that when data collection begins, the mean of all runs at time $t > T_W$ (blue solid line) has unbiased fluctuations about the time-averaged mean (red solid line). The equilibrium position $h_\mathrm{eq}$ is taken as the time-independent mean, with the standard error as the statistical uncertainty (red shading). 

\subsection{Methods for attraction and repulsion zones}
\label{sctn:attRepZones}

\begin{figure*}[tb]
    \centering
    \includegraphics[width=1\linewidth]{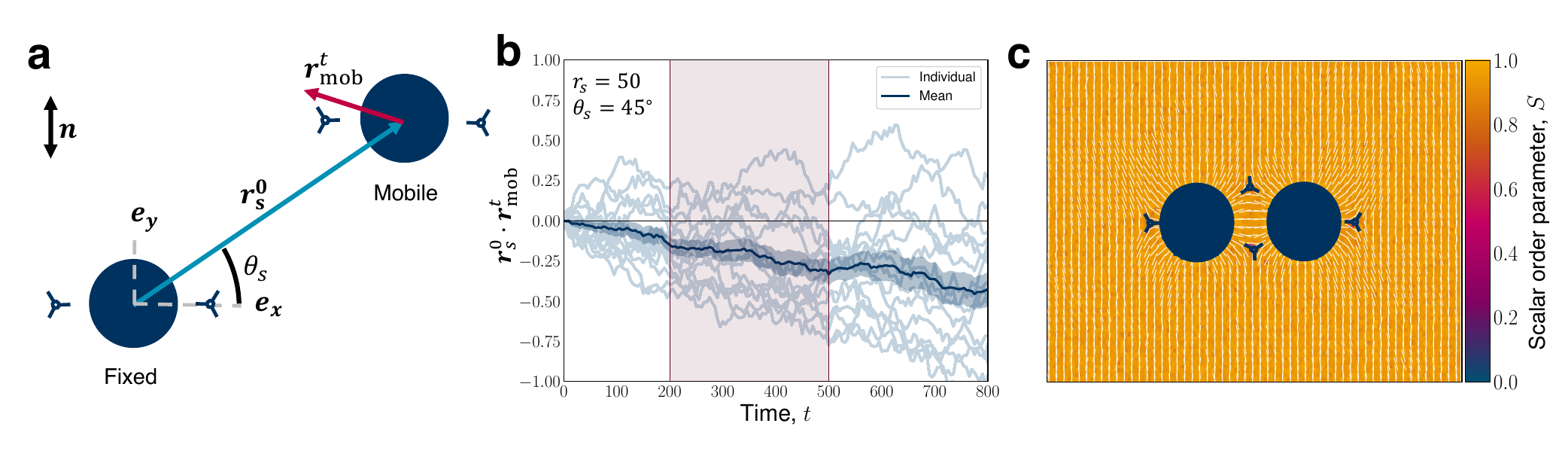}
    \caption{Measurements for determining the two-dimensional attraction-repulsion between a fixed and mobile colloid. 
    \textbf{(a)} Schematic with colloids (navy circles) initially separated by $\vec{r}_s^0$ with polar angle $\theta_s$ relative to $\basis_x$. 
    The far-field director is initialised along $\dir$, which, at short times, fixes the orientation of the $-1/2$ defects (blue trefoil symbols) to the colloids' equators. 
    At time $t$, the mobile colloid moves to a new position with displacement vector $\vec{r}_{\text{mob}}^t$. 
    \textbf{(b)} Example trajectories for $15$ simulations (grey) at a starting separation $r_s^0=50$ and angle $\theta_s=45^\circ$. 
    The trajectories are measured as the projection magnitude of the mobile displacement onto the axis of the initial separation $\mathbf{r}^0_s\cdot \mathbf{r}_{\mathrm{mob}}^t$. 
    The attraction-repulsion behaviour is measured over the time interval $t=200$ to $t=500$, as indicated by the red shaded region. 
    The mean is shown as the solid dark blue line, and standard error as blue shading.
    \textbf{(c)} Snapshot of self-assembly into a chain. 
    Defects are shared between the colloidal disks, influencing an attractive response at small angles and separations.
    }
    \label{fig:ARmethods}
\end{figure*}

The force anisotropy measurements are obtained in two-dimensions for simplicity. 
The angular dependence of the interaction between two colloids is determined by fixing one colloid, placing a mobile probe colloid in its vicinity and measuring the response of the probe colloid (\fig{fig:ARmethods}a). 
The far-field director alignment is initialised along $\basis_y$, which preferentially positions two $-1/2$ defects on either side of the colloids, establishing consistent initial quadrupole orientations.
A short warmup period of $T_W=5$ allows the defects to form but not reorient away from alignment in $\basis_y$. 
The two colloids are placed, one fixed at $(40,40)$ and the other mobile initialised from $(40+\Delta x,40 + \Delta y)$, where $\Delta x = r_s \cos \theta_s$ and $\Delta y = r_s \sin \theta_s$ with separation magnitude $r_s\in [40,50,55,60]$ and orientation angle, relative to $\basis_x$, of $\theta_s\in [0^\circ,6^\circ,11^\circ,17^\circ,22^\circ,33^\circ,45^\circ,56^\circ,68^\circ,79^\circ,90^\circ]$. Simulations have periodic boundary conditions on all walls, with system size $L_x=80+40\cos\theta_s$ and $L_y=80+40\sin\theta_s$. The simulation time is $T_S=3000$.

Two vectors are measured to determine the attractive or repulsive behaviour of the mobile colloidal probe placed at varying separations $r_s$ and angular positions $\theta_s$ relative to a fixed colloid (\fig{fig:ARmethods}a).
The first is the initial separation vector between the colloids $\mathbf{r}_s^0=\mathbf{r}_2^0-\mathbf{r}_1^0$ (blue arrow). This establishes a constant reference to measure the response of the mobile colloid. The second $\mathbf{r}_{\mathrm{mob}}^t=\mathbf{r}_2^t-\mathbf{r}_2^0$ is a temporally varying separation vector, which records the displacement of the mobile colloid at time $t$ from the start position (red arrow). 
Individual mobile colloids are regarded to have repulsive or attractive behaviour if the projection of $\mathbf{r}_{\mathrm{mob}}^t$ on $\mathbf{r}_s^0$ is positive or negative respectively.
The individual trajectories are noisy (grey trajectories in \fig{fig:ARmethods}b) and, after some time, the $-1/2$ defects reorient to aid self-assembly into chains (\fig{fig:ARmethods}c). This reorientation misaligns the relative quadrupole orientations.  
Therefore, $N=15$ simulation runs are performed for each combination of $r_s$ and $\theta_s$, and the response behaviours are measured from the early time dynamics chosen to be $ 200 \leq t \leq 500$. The minimum time of $t = 200$ is chosen to establish sufficient statistical certainty on the attraction-repulsion trajectories.
Ensemble averages of the projection magnitude $\vec{r}_s^0 \cdot \vec{r}_{\mathrm{mob}}^t$ are performed, extracting the mean $\mu$ and standard error $\sigma_M=\sigma/\sqrt{N}$. The nature of each colloidal site ($r_s\cos\theta_s,r_s\sin\theta_s$) is calculated as attractive if $\mu \leq -\sigma_M$, repulsive if $\mu \geq \sigma_M$ and neutral otherwise. 

\subsection{Defect analysis}
\label{sctn:defects}

Disclination loops are identified using the disclination density tensor, proposed by Schimmings and Vinals \cite{schimming2022}. Using Einstein-index summation convention for clarity, the tensor is conveniently constructed from derivatives of the nematic $\Qtens-$tensor
\begin{align}
    D_{ij} &= \epsilon_{i\mu\nu}\epsilon_{jlk} \partial_l Q_{\mu\alpha} \partial_k Q_{\nu\alpha},
    \label{eq:D}
\end{align}
where $i,j,k,\alpha,\mu,\nu$ are tensor indices corresponding to $\basis_{x},\basis_{y},\basis_{z}$. The disclination density tensor $\tens{D}$ can be directly interpreted as the dyad
\begin{align}
    \label{eq:Dtensordecomposition}
    \tens{D} &= s(\pos) \rotVecdG \otimes \tanVecdG ,
\end{align}
composed of the tangent vector $\tanVecdG$ of the disclination line and the rotation vector $\rotVecdG$, which defines the winding plane of the director in the vicinity of the disclination \cite{friedel1969}. 
The relative angle between them $\cos\beta=\rotVecdG\cdot\tanVecdG$ illustrates if the local disclination has a wedge profile (with $\cos\beta = +1$ for $+1/2$ defect profiles and $\cos\beta = -1$ for $-1/2$ defect profiles) or a twist profile (with $\cos\beta \approx 0$). 

The scalar field $s(\pos)$ is non-negative, and is maximum at the core of the disclination -- therefore providing a useful quantity for identifying disclinations, with an appropriately defined lower bound. 
Throughout this study, disclinations are identified as $s(\pos)\geq 0.9$, which was found to produce smoother disclinations than using isosurfaces of the nematic scalar order parameter $S(\pos)$. 
Extracting $s(\pos)$, $\rotVecdG$ and $\tanVecdG$ from \eq{eq:Dtensordecomposition} utilises the methods outlined in \cite{schimming2022}. The vectors $\rotVecdG$ and $\tanVecdG$ are ensured to be continuous and have the correct relative sign by: 1) applying a clustering algorithm that groups disclination cells into disclination lines, 2) ensuring the tangent vector smoothly varies along the line and 3) fixing the sign using $\sgn{\rotVecdG\cdot\tanVecdG} = \sgn{ \tr{\tens{D}} }$. 

A second clustering algorithm groups disclination cells into an ordered sequence of larger points, that combines together a group of nearest neighbours, without reusing cells from other groups. The start and end point of the sequence connect together to form a loop. The $\tanVecdG$ and $\rotVecdG$ of composite cells are averaged over to return a single dyad per point. This construction into points enables geometric properties of the loop to be established, particularly those required to calculate the ribbon properties of the disclination line.

\subsection{Ribbon framing}
\label{sctn:ribbon}

To construct the ribbon, a framing vector $\framing$ perpendicular to $\tanVecdG$ is required that varies continuously along the disclination loop.  
In the proximity of the disclination, $\dir$ is oriented in a plane with normal vector $\rotVecdG$, which is anti-parallel to $\tanVecdG$ as confirmed by the colouring $\rotVecdG\cdot\tanVecdG = \cos\beta=-1$. We therefore make an arbitrary choice to track one of the three radially pointing orientations of the $-1/2$ profile along the disclination.
To extract the three radial orientations, $\framing_1,\framing_2,\framing_3$, we construct a small cube of $5\times5\times5$ lattice cells centred on each of the ordered points. For each of these cubes, a rotation matrix $\mathcal{R}$ is constructed and applied to the director field within the cube, which aligns the local disclination tangent $\tanVecdG$ with $\basis_z$. This enables the director radial orientation to be identified on the transformed $\basis_x\basis_y$ plane, on which we construct a test vector $\vec{r}_{\mathrm{test}}=(\cos\theta_{\mathrm{test}},\sin\theta_{\mathrm{test}})$ oriented radially outwards from the core. On a circuit of points  $\pos_l$ surrounding the core, $\vec{r}_{\mathrm{test}}$ is compared with the local director $\dir(\pos_l)$. Determined over all points $\pos_l$ and test orientations $\theta_{\mathrm{test}}$, $\framing_1$ is chosen as the $\vec{r}_{\mathrm{test}}$ that maximises the absolute value $|\vec{r}_{\mathrm{test}}\cdot\dir(\pos_l)|$. The inverse rotation transform $\mathcal{R}^{-1}$ is applied to $\framing_1$ to revert back to the original basis, and $\framing_2,\framing_3$ are determined as orientations $2\pi/3$ rotated relative to each other about $\tanVecdG$.
The framing vector is initialised as $\framing=\framing_1$ for the first point along the loop, and subsequent points choose from one of the three $\framing_1,\framing_2,\framing_3$ orientations that minimise the rotation angle compared with $\framing$ from the previous point in the sequence.

\subsection{Calculating the self-linking number}
\label{sctn:self-linking}

The self-linking number is calculated through the geometric writhe $\Wr$ and twist $\Tw$ properties of the disclination via \eq{eq:sl}.
Twisting is the local winding of the framing vector around the tangent curve, which gives $\Tw$ when integrated. Writhe is a non-local geometric property that describes the coiling of the curve, through tracking the relative rotation of locally parallel tangent bundles along the loop \cite{copar2011n2}. 
Writhe and twist are calculated as 
\begin{align}
    \Wr &= \frac{1}{4\pi}\oint_C ds \oint_C ds' \ 
        \tanVecdG(s) \times \tanVecdG(s')\cdot \frac{\vec{R}(s)-\vec{R}(s')}{|\vec{R}(s)-\vec{R}(s')|^3} \\
    \Tw &= \frac{1}{2\pi}\oint_C ds \ 
        \tanVecdG(s) \cdot \left(\framing(s) \times \frac{d \framing(s)}{ds}\right) ,
\end{align}
where $\vec{R}(s)$ are position vectors for points along the loop, $\tanVecdG(s)=\frac{d \vec{R}(s)}{ds}$ is the local tangent vector~\cite{Kamien2002}, and $C$ is a closed curve composing the disclination loop. 
The framing vector $\framing(s)$ is everywhere perpendicular to $\tanVecdG(s)$ and sets up the local framing direction (\sctn{sctn:ribbon}).

\balance

\bibliography{main}

\bibliographystyle{unsrt}

\end{document}